\documentclass[a4paper,11pt]{article}
\pdfoutput=1 	% if your are submitting a pdflatex (i.e. if you have
\usepackage{jheppub} 	% for details on the use of the package, please
\usepackage[utf8]{inputenc}
\usepackage[T1]{fontenc}
\usepackage{lmodern}
\usepackage{amsmath, amssymb}
\usepackage{microtype}  % improves the typesetting in general
\usepackage{graphicx}	% Include figure files
\usepackage{dcolumn}	% Align table columns on decimal point
\usepackage{bm}			% bold math
\usepackage{hyperref}	% add hypertext capabilities
\usepackage{color} 		% colored text
\usepackage{tabularx}   % control table width
\usepackage{orcidlink}  % Orcid IDs

\definecolor{ao}{rgb}{0.0, 0.5, 0.0}

\newcommand{\GeV}{\,\mathrm{GeV}}
\newcommand{\MeV}{\,\mathrm{MeV}}
\newcommand{\keV}{\,\mathrm{keV}}
\newcommand{\eV}{\,\mathrm{eV}}

\title{Novel constraints on neutrino physics beyond the standard model from the CONUS experiment}% Force line breaks with \\
\author[a]{H.~Bonet,}
\author[a]{A.~Bonhomme\,\orcidlink{0000-0002-0218-2835},}
\author[a]{C.~Buck\,\orcidlink{0000-0002-5751-5289},}
\author[b]{K.~F\"ulber,}
\author[a]{J.~Hakenm\"uller\,\orcidlink{0000-0003-0470-3320},}
\author[a]{G.~Heusser,}
\author[a]{T.~Hugle\,\orcidlink{0000-0001-9788-4014},}
\author[a]{M.~Lindner\,\orcidlink{0000-0002-3704-6016},}
\author[a]{W.~Maneschg\,\orcidlink{0000-0003-0320-7827},}
\author[a]{T.~Rink\,\orcidlink{0000-0002-9293-1106},}
\author[a]{H.~Strecker}
\author[b]{and R.~Wink}

\affiliation[a]{Max-Planck-Institut f\"ur Kernphysik, Saupfercheckweg 1, 69117 Heidelberg, Germany}
\affiliation[b]{PreussenElektra GmbH, Kernkraftwerk Brokdorf, Osterende, 25576 Brokdorf, Germany}

\collaboration{CONUS Collaboration}
\emailAdd{conus.eb@mpi-hd.mpg.de}

\date{\today}% It is always \today, today,
\abstract{
The measurements of coherent elastic neutrino-nucleus scattering (CE$\nu$NS) experiments have opened up the possibility to constrain neutrino physics beyond the standard model of elementary particle physics.
Furthermore, by considering neutrino-electron scattering in the keV-energy region, it is possible to set additional limits on new physics processes.
Here, we present constraints that are derived from \textsc{Conus} germanium data on beyond the standard model (BSM) processes like tensor and vector non-standard interactions (NSIs) in the neutrino-quark sector, as well as light vector and scalar mediators.
Thanks to the realized low background levels in the \textsc{Conus} experiment at ionization energies below $1 \keV$, we are able to set the world's best limits on tensor NSIs from CE$\nu$NS and constrain the scale of corresponding new physics to lie above $360\GeV$.
For vector NSIs, the derived limits strongly depend on the assumed ionization quenching factor within the detector material, since small quenching factors largely suppress potential signals for both, the expected standard model CE$\nu$NS process and the vector NSIs.
Furthermore, competitive limits on scalar and vector mediators are obtained from the CE$\nu$NS channel at reactor-site which allow to probe coupling constants as low as $5\cdot10^{-5}$ of low mediator masses, assuming the currently favored quenching factor regime.
The consideration of neutrino-electron scatterings allows to set even stronger constraints for mediator masses below $\sim1\MeV$ and $\sim 10\MeV$ for scalar and vector mediators, respectively.
}

\begin{document}
\maketitle
\flushbottom

%%%%%%%%%%%%%%%%%%%%%%%%%%%%%%%%%%%%%%%%%%%%%%%%%%%%%%%%%%%%%%%%%%%%%%%%%%%%%%%%
%%%%%%%%%%%%%%%%%%%%%%%%%%%%%%%%%%%%%%%%%%%%%%%%%%%%%%%%%%%%%%%%%%%%%%%%%%%%%%%%

\section{Introduction}\label{sec:intro}

Coherent elastic neutrino-nucleus scattering (CE$\nu$NS) is a standard model (SM) process of elementary particle physics that was predicted shortly after the discovery of the $Z$-boson~\cite{Freedman:1973yd,Tubbs:1975jx,Drukier:1983gj}.
After over forty years, first observations of this process were reported by the \textsc{Coherent} Collaboration, using a pion-decay-at-rest ($\pi$DAR) source in combination with scintillation and liquid noble gas detectors~\cite{COHERENT:2017ipa,COHERENT:2020iec}.
The \textsc{Conus} experiment pursues detecting this interaction channel with reactor electron antineutrinos and recently published first limits~\cite{CONUS:2020skt}.
The underlying data were acquired with low background germanium detectors located at 17.1\,m distance from the reactor core center of the 3.9\,GW (thermal power) nuclear power plant in Brokdorf, Germany.

So far, no deviations from the SM prediction have been observed in the operational experiments. 
However, new possibilities to search for physics beyond the standard model (BSM) have already triggered various phenomenological investigations~\cite{Kosmas:2017tsq,Billard:2018jnl,Khan:2019cvi,Papoulias:2019txv}.
Together with their expected SM interactions, any new interaction of neutrinos can play an important role in a wide range of physics branches: from cosmology to the smallest scales of nuclear and particle physics.
In an astronomical context, they play a key role in the evolution of stellar collapses~\cite{Freedman:1977xn,Amanik:2006ad} and might influence stellar nucleosynthesis~\cite{Balasi:2015dba}.
In addition, with neutrino detection via CE$\nu$NS at hand, flavor-independent astronomy with supernova neutrinos becomes feasible~\cite{Amaya:2011sn,Brdar:2018zds,Agnes:2020pbw} and thus allows to investigate the interior of dense objects as well as stellar evolution in detail.
The next-generation dark matter direct-detection experiments will face an irreducible background, the so-called neutrino-floor, which is caused by atmospheric, solar and supernova remnant neutrinos that coherently scatter in such detectors~\cite{Monroe:2007xp,Gutlein:2010tq}.
From the perspective of neutrino physics, this opens up new possibilities as new neutrino interactions might manifest themselves in this ``background'' as well~\cite{Cerdeno:2016sfi,Bertuzzo:2017tuf,Boehm:2018sux,Link:2019pbm,AristizabalSierra:2019ykk}.

In a nuclear and particle physics context, even without any new physics contributions, CE$\nu$NS can allow for a determination of the neutron density distribution of a target nucleus~\cite{Patton:2012jr,Cadeddu:2017etk,Coloma:2020nhf,VanDessel:2020epd} as well as the weak mixing angle in the unexplored MeV regime~\cite{Lee:2015yht,Canas:2018rng,Huang:2019ene,Fernandez-Moroni:2020yyl}.

For BSM searches, CE$\nu$NS detectors can be used to search for non-standard neutrino-quark interactions (NSIs)~\cite{Barranco:2005yy,Barranco:2007tz,Barranco:2011wx,Lindner:2016wff,Coloma:2017ncl,Liao:2017uzy,Bischer:2018zbd,Dev:2019anc,Giunti:2019xpr,Denton:2020hop} and to investigate potential electromagnetic properties of the neutrino~\cite{Vogel:1989iv,Giunti:2014ixa,Cadeddu:2018dux,Miranda:2019wdy,Cadeddu:2020lky}, e.g.\ finite magnetic moments or a millicharge.
Being at lower energy scales than typical collider experiments, CE$\nu$NS experiments complement their BSM searches and might result in either competitive or even stronger bounds for light mediators~\cite{deNiverville:2015mwa,Dent:2016wcr}.
In particular, investigations of light scalars and/or axion-like particles~\cite{Farzan:2018gtr,Dent:2019ueq,AristizabalSierra:2020rom}, and light vectors~\cite{Dutta:2019eml,Aguilar-Arevalo:2019zme,Cadeddu:2020nbr,Miranda:2020zji}, e.g. dark photons, take advantage of this new channel.
Even searches for new fermions seem possible within the context of CE$\nu$NS measurements~\cite{Brdar:2018qqj,Chang:2020jwl}.

More generally, a high statistics CE$\nu$NS measurement can be used to determine the flux of a neutrino source precisely. 
Regarding the flux anomalies reported from several short-baseline experiments and the possible eV-mass sterile neutrino solution~\cite{Aguilar:2001ty,Mention:2011rk,Boser:2019rta}, CE$\nu$NS might contribute further knowledge, especially since it provides flavor-blind and energy-threshold-free information about the source's (anti)neutrino spectrum~\cite{Formaggio:2011jt,Dutta:2015nlo,Canas:2017umu,Blanco:2019vyp,Miranda:2020syh}.
Particularly at nuclear reactors, small (and therefore simpler to integrate) CE$\nu$NS sensitive devices could help in monitoring their power and flux and, in the future, even determine a reactor's antineutrino spectrum below $1.8 \MeV$, which is usually limited by the threshold energy of the used detection channel, i.e.\ inverse beta-decay (IBD).
In this way, neutrino physics might help in reactor safeguarding and contribute to nuclear non-proliferation~\cite{Hagmann:2004uv,Bernstein:2019hix,Bowen:2020unj}.

All the above mentioned SM and BSM possibilities in combination with improvements in detector and background suppression techniques have made CE$\nu$NS measurements a feasible and promising endeavor both at neutrino $\pi$DAR sources and nuclear reactors.
While the \textsc{Coherent} Collaboration is preparing the operation of further detector systems with different target elements at a $\pi$DAR neutrino source, there are many more experimental attempts to measure CE$\nu$NS with electron antineutrinos emitted from nuclear reactors: \textsc{Connie}~\cite{Aguilar-Arevalo:2016khx}, \textsc{Miner}~\cite{Agnolet:2016zir}, \textsc{Ncc-1701} at \textsc{Dresden-II}~\cite{Colaresi:2021kus}, \textsc{Neon}~\cite{Choi:2020gkm}, $\nu$-cleus~\cite{Strauss:2017cuu}, $\nu$\textsc{Gen}~\cite{Belov:2015ufh}, \textsc{Red}-100~\cite{Akimov:2019ogx}, \textsc{Ricochet}~\cite{Billard:2016giu} and \textsc{Texono}~\cite{Wong:2004ru}.
In these reactor experiments, different detection technologies are used, e.g.\ charged-coupled devices (CCDs)~\cite{FernandezMoroni:2014qlq}, cryogenic calorimeters~\cite{Strauss:2017cam}, high-purity germanium (HPGe) crystals~\cite{Bonet:2020ntx}, liquid noble gas detectors~\cite{Chepel:2012sj} as well as scintillating crystals~\cite{Choi:2020qcj}.
In this way, the field of CE$\nu$NS is going to be probed with the full range of recent detector technologies and different target nuclei -- each with its own particular advantages and complementarities -- allowing to expect interesting results from SM as well as BSM investigations.

As a part of the experimental efforts in this direction, we present here the first BSM results derived from the \textsc{Conus} \textsc{Run}-1 data.
We use a very similar analysis procedure to the one employed for the experiment's first CE$\nu$NS limit determination~\cite{CONUS:2020skt} and apply it to common BSM models that have already been investigated in the context of other CE$\nu$NS measurements.
In particular, we show bounds on tensor and vector NSIs as well as simplified light vector and scalar mediator models.
For the latter two, we deduce bounds from neutrino scattering off electrons and off nuclei.

\noindent This paper is structured as follows:
In Section~\ref{sec:methods} we describe the analysis method that is used for the BSM models in the course of this paper.
Next to a general introduction of the \textsc{Conus} set-up, we give an overview of the analysis procedure as well as systematic uncertainties that underlie this investigation.
We further introduce two data sets that are chosen for the two scattering channels under study, i.e.\ neutrino-nucleus and neutrino-electron scattering.
Subsequently, we show the results of the performed investigations in Section~\ref{sec:bsm_study}.
Limits on tensor and vector NSIs are presented and in the context of light vector and scalar mediator searches, we derive bounds from electron scattering in the ionization energy region between $2$ and $8 \keV_{ee}$.\footnote{
The notations ``eV$_{ee}$'' and ``eV$_{nr}$'', will be used in the following as a shorthand notation to distinguish ionization energy, denoted as $ee$ (as a reference to ``electron equivalents''), and nuclear recoil energy, denoted as $nr$.}
Finally, in Section~\ref{sec:conclusion} we conclude and give an outlook on the various BSM investigations that will become feasible with \textsc{Conus} and the next generation of CE$\nu$NS experiments.

%%%%%%%%%%%%%%%%%%%%%%%%%%%%%%%%%%%%%%%%%%%%%%%%%%%%%%%%%%%%%%%%%%%%%%%%%%%%%%%%
%%%%%%%%%%%%%%%%%%%%%%%%%%%%%%%%%%%%%%%%%%%%%%%%%%%%%%%%%%%%%%%%%%%%%%%%%%%%%%%%

\section{Data sets, experimental framework and analysis method}
\label{sec:methods}
For the analysis presented here, we use the \textsc{Conus} \textsc{Run}-1 data and employ a binned likelihood analysis to derive limits on parameters of the considered BSM models.
In addition to the \textsc{Run}-1 data set used for the CE$\nu$NS analysis described in Ref.~\cite{CONUS:2020skt}, we work with a second \textsc{Run}-1 data set at energies between $2$ and $8 \keV_{ee}$, which exhibits longer data collection periods for the BSM channels that are sensitive to neutrino-electron scattering.
The details of both data sets as well as the likelihood analysis are laid out in the following subsections.

%%%%%%%%%%%%%%%%%%%%%%%%%%%%%%%%%%%%%%%%%%%%%%%%%%%%%%%%%%%%%%%%%%%%%%%%%%%%%%%%

\subsection{Data sets and the experimental framework of the CONUS experiment}
\label{subsec:datasets_experimentalframework}
The data sets used in this BSM analysis were gathered during \textsc{Run}-1 (Apr 01 - Oct 29, 2018) of the \textsc{Conus} experiment which is operated at the commercial nuclear power plant in Brokdorf, Germany.
Inside the nuclear power plant is a single-unit pressurized water reactor that is operated at a maximal thermal power of 3.9\,GW and serves as an intense electron antineutrino source at the 17m-distant experimental site.
The expected antineutrino spectrum is a typical reactor spectrum, dominated by the contribution of the four isotopes $^{235}$U, $^{238}$U, $^{239}$Pu and $^{241}$Pu~\cite{Hayes:2016qnu}, with all of the neutrinos having energies of less than $\sim\! 10 \MeV$.
To describe the antineutrino emission spectrum from the reactor, we start from the predicted antineutrino spectra by Huber and Müller~\cite{Huber:2011wv,Mueller:2011nm} and correct for the $5 \MeV$-bump observed in experimental data~\cite{An:2016srz}.
The relative contribution of the different isotopes can be accounted for by weighting the different isotopes according to their time-dependent fission fractions, which are provided to us by the reactor operating company PreussenElektra GmbH.
The corresponding values for the three detectors \textsc{Conus}-1, \textsc{Conus}-2 and \textsc{Conus}-3 (C1-C3) considered in the following analyses are listed in Table~\ref{tab:fission_fractions}.
%
%%%%%%%%%%%%%%%%%%%%%%%%%%%%%%%%%%%%%%%%%%%%%%%%%%%%%%%%%%%%%%%%%%%%%%%%%%%%%%%%
{\renewcommand{\arraystretch}{1.5}%
\setlength{\tabcolsep}{0.45cm}%
\begin{table}[t]
    \centering
    \begin{tabular}{c | c c c c | c }
      Detector & $^{235}$U [\%] & $^{238}$U [\%] & $^{239}$Pu [\%] & $^{241}$Pu [\%] & $\bar{P}_{\mathrm{th}}$ [\%]\\
      \hline
      C1 & 60.3; 56.8 & 7.1; 7.2 & 27.0; 29.9 & 5.4; 6.1 & 92.33; 89.88 \\
      C2 & 63.8; 56.9 & 7.1; 7.2 & 24.2; 29.8 & 4.9; 6.1 & 92.70; 90.12 \\
      C3 & 57.2; 56.8 & 7.2; 7.2 & 29.7; 29.9 & 6.0; 6.1 & 88.79; 90.10 \\
    \end{tabular}
    \caption{Average fission fractions of the most relevant isotopes in the reactor antineutrino spectrum and average reactor powers $\bar{P}_{\mathrm{th}}$ in terms of the reactor's maximal thermal power of $3.9$\,GW for standard/extended data sets of \textsc{Conus} \textsc{Run}-1.
    The detectors C1-C3 used in the following analyses are assigned individual values due to their specific data collection periods.}
    \label{tab:fission_fractions}
\end{table}}
%
%%%%%%%%%%%%%%%%%%%%%%%%%%%%%%%%%%%%%%%%%%%%%%%%%%%%%%%%%%%%%%%%%%%%%%%%%%%%%%%%
%
This reactor spectrum above the $1.8 \MeV$ threshold of IBD experiments determines the neutrino spectrum for all processes associated with nucleus scattering.
For the electron scattering channels that we analyze, also the low-energy part (below $1.8 \MeV$) of the spectrum becomes relevant for which we use the simulation data provided by Ref.~\cite{Kopeikin:2003gu}.
These simulations for the different isotopes can be weighted by the fission fractions and normalized to the total number of neutrinos emitted over the whole spectrum, of which there are on average $\sim \! 7.2$ per fission, cf.~Ref.~\cite{Beda:2007hf}.
To determine the total flux of antineutrinos that can interact with the \textsc{Conus} detectors, we can use the total number of fissions per second derived from the reactor thermal power, as every fission releases about $200 \MeV$ of energy (cf.~Ref.~\cite{Ma:2012bm} for details and exact isotope specific values).
This leads to a total antineutrino flux at the experimental site of $2.3 \cdot 10^{13}$\,s$^{-1}$\,cm$^{-2}$.
The influence of the shape uncertainties, i.e.\ the covariance matrix of the neutrino spectrum as provided by Ref.~\cite{An:2016srz}, was investigated in the context of the \textsc{Conus} CE$\nu$NS analysis~\cite{CONUS:2020skt} and turned out to be negligible in our case.
Therefore, we do not include them in the present analysis.

%%%%%%%%%%%%%%%%%%%%%%%%%%%%%%%%%%%%%%%%%%%%%%%%%%%%%%%%%%%%%%%%%%%%%%%%%%%%%%%%
\begin{figure}
	\centering
	\includegraphics[width=\textwidth]{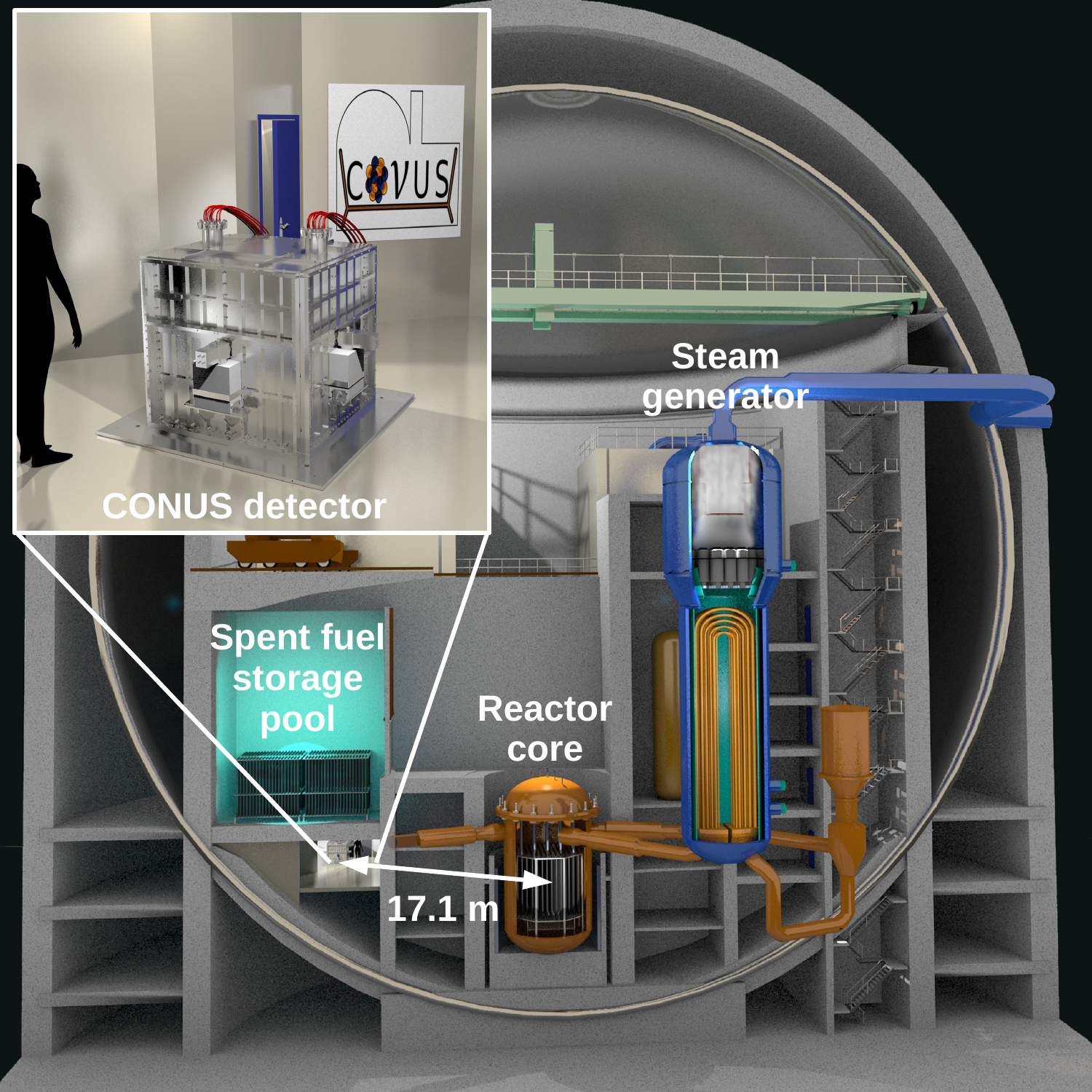}
	\caption{
	Position of the \textsc{Conus} detector set-up within the building of the nuclear power plant at Brokdorf, Germany.
	It is located under the spent fuel storage pool at 17.1\,m distance to the 3.9\,GW (thermal power) reactor core. 
	The vertical position of the set-up coincides approximately with the reactor core's center.
	The enlarged image shows the set-up at its experimental site.
	Within the shown stainless steel cage, layers of lead as well as pure and borated polyethylen serve as passive shield around the embedded four HPGe detectors against external radiation and other background sources. 
	Further, it includes plastic scintillator plates equipped with photomultiplier tubes which are used as muon veto.
	}
	\label{fig:conus_kbr}
\end{figure}
%%%%%%%%%%%%%%%%%%%%%%%%%%%%%%%%%%%%%%%%%%%%%%%%%%%%%%%%%%%%%%%%%%%%%%%%%%%%%%%%

\noindent Besides the immense reactor flux and the corresponding spectral distribution of antineutrinos, the achieved background level with the deployed shield is another cornerstone of the whole experimental framework.
The shield is extremely compact, with a volume of only 1.65\,m$^3$ and a mass of 11\,tonnes, and exhibits an onion-like structure.
It consists of lead bricks, borated and non-borated polyethylene plates, and plastic scintillator plates equipped with photomultiplier tubes serving as an active muon anticoincidence system (muon veto).
Around the layers, a protective stainless steel cage helps fulfilling the safety requirements.
The shield design is based on the long-time experience with low background techniques at Max-Planck-Institut für Kernphysik (MPIK), e.g.~Refs.~\cite{Heusser:1995wd,Heusser:2015ifa}, while being optimized to the experimental site at shallow depth next to a reactor core.
The location of the CONUS detector and the dimension of the whole set-up within the nuclear power plant are illustrated in Figure~\ref{fig:conus_kbr}.

The influence of possible reactor-correlated background types was confirmed to be negligible via dedicated neutron and $\gamma$-ray measurement campaigns.
These were supported by validated background Monte Carlo (MC) simulations that incorporated a large fraction of the reactor geometry surrounding the experimental site~\cite{Hakenmuller:2019ecb}.
Thus, the background to the BSM analyses is uncorrelated to the reactor thermal power. 
It is described like in the CE$\nu$NS investigation by MC simulations.
For the BSM analyses of both scattering channels, the background model is almost identical to the one used in the CE$\nu$NS publication, cf.~Ref.~\cite{CONUS:2020skt}. 
Only small adjustments to the background model have been made for the extended data sets, which are used for the electron scattering channels.
In that context due to the extended region of interest (ROI) to higher energies, systematic uncertainties on the spectral shape of the background model are considered in order to account for uncertainties regarding the production rate of cosmogenic induced isotopes as well as surface contamination on the Ge diodes. 
Details of the applied background model and its uncertainties can be found in a dedicated background description of the \textsc{Conus} experiment, cf.~Ref.~\cite{CONUSbackground}.
In the energy window of $500$ to $1000 \eV_{ee}$, just above the ROI for CE$\nu$NS studies, the \textsc{Conus} detectors achieve background levels of a few $10\, \mathrm{\, counts \, kg^{-1} \, d^{-1} \, keV}_{ee}^{-1}$, while having an effective overburden of 24\,m of water-equivalent (m w.e.) only.

To detect the antineutrinos that cross the shield, \textsc{Conus} uses four 1\,kg-sized point-contact HPGe spectrometers with sub-keV$_{ee}$ energy thresholds. 
A full description can be found in Ref.~\cite{Bonet:2020ntx}.
The four detectors have a total active mass of $(3.73\, \pm \, 0.02)\,$kg and provide the necessary characteristics for a CE$\nu$NS measurement at a commercial reactor site: ultra-low noise levels and thus very low energy thresholds, i.e.\ $\lesssim300\eV_{ee}$, low concentrations of radioactive contamination as well as electrically powered cryocoolers.
Within a CE$\nu$NS process, the induced nuclear recoil releases heat and ionization electrons that might be collected by an appropriate detector for signal formation.
However, in the present case, only the ionization energy part is registered by the HPGe detectors, resulting in an energy that is suppressed by $75-85\%$ compared to the original recoil energy.
This phenomenon is commonly referred to as `quenching'.
Consequently, this makes detecting CE$\nu$NS signals even more difficult.
To take the effect of quenching into account, we apply the widely used Lindhard model~\cite{Lindhard:1961zz}, modified with an adiabatic correction~\cite{Scholz:2016qos}.
Its associated parameter $k$ roughly corresponds to the quenching factor at nuclear recoils of $\sim1 \keV_{nr}$.
One recent measurement indicates that quenching deviates from this description especially at ionization energies of $\sim \! 250 \eV_{ee}$ and below, cf.~Ref.~\cite{Collar:2021fcl}.\footnote{Note that for such low energies, simplifying assumptions underlying the Lindhard model can be questioned and deviations might be described by an additional parameter~\cite{Sorensen:2014sla}.}
Thus, an accurate determination of the quenching factor cannot only support CE$\nu$NS measurements, but also affects BSM studies~\cite{Liao:2021yog} as it appears in any process that involves scattering off a nucleus.
So far, there is a variety of measurements for the quenching factor in germanium with larger systematic uncertainties that still leave enough room to constitute the dominating source of uncertainty for our BSM analyses here.
To account for this uncertainty, we always present the results for different quenching factors which cover the range of currently available experimental data.

Generally, the \textsc{Conus} data acquisition is divided into reactor \textsc{On} and reactor \textsc{Off} periods as well as periods reserved for commissioning and optimization.
Each data set then has been defined individually according to the stability of environmental parameters like ambient temperature.
For the details of this data selection procedure we refer to Ref.~\cite{Bonet:2020ntx}.
In the present analysis, we use data of the first acquisition period which we refer to as \textsc{Run}-1 data set.
For this data set, the \textsc{Conus}-4 (C4) detector is excluded due to a temporarily appearing  artifact, cf.~Ref.~\cite{CONUS:2020skt}.
Besides neutrino-nucleus scattering, where only the region below $1 \keV_{ee}$ is important, we also analyze neutrino-electron scattering at energies between $2$ and $8 \keV_{ee}$.
We limit our analysis of the electron channel to this energy interval because of two reasons:
First, we are looking at signals that emerge as broader spectral contribution above the continuum of the spectrum. 
The selected region is line-free and naturally confined by x-ray peaks around $\sim1\keV$ and $\sim10\keV$, which are due to K- and L-shell transitions in decays of Ge-related isotopes. 
These isotopes were/are produced by cosmic activation above ground and partially in-situ at the experimental site, as well as via sporadically deployed artificial neutron calibration sources.
Second, the new ROI is not affected at all by potential noise, that is correlated with the ambient temperature, cf.~Ref.~\cite{Bonet:2020ntx}, and which caused an exclusion of parts of the data from our first CE$\nu$NS analysis in the sub-keV regime. 
Thereby, we can increase the total lifetime of the extended data set, compared to the CE$\nu$NS data set, by a factor of 3.1 for \textsc{On} and a factor of 2.5 for \textsc{Off} periods.
The specifications of all final data sets after data selection and cuts, used for the BSM analysis in this paper, are depicted in Table~\ref{tab:datasets}.

%%%%%%%%%%%%%%%%%%%%%%%%%%%%%%%%%%%%%%%%%%%%%%%%%%%%%%%%%%%%%%%%%%%%%%%%%%%%%%%%

{\renewcommand{\arraystretch}{1.2}%
\setlength{\tabcolsep}{0.45cm}%
\begin{table}[t]
    \centering
    \begin{tabular}{c | c c c c }
      Scattering channel & Detector & \textsc{On} [kg\,d] & \textsc{Off} [kg\,d] & ROI [eV$_{ee}$]\\ 
      \hline
      & C1 & 96.7 & 13.8 & 276 - 741\\
      $\bar{\nu}_e + A(Z, N)$ & C2 & 14.6 & 13.4 & 281 - 999\\
      & C3 & 97.5 & 10.4 & 333 - 991\\
      %\hline
      & all & 208.8 & 37.6 & \\
      \hline
      & C1 & 215.4 & 29.6 & 2013 - 7968\\
      $\bar{\nu}_e + e$ & C2 & 184.6 & 32.2 & 2006 - 7990\\
      & C3 & 248.5 & 31.7 & 2035 - 7989\\
      & all & 648.5 & 93.5
    \end{tabular}
    \caption{Lifetimes for reactor \textsc{On} and \textsc{Off} periods together with the regions of interest (ROIs) for the different detectors in both scattering channels during \textsc{Run}-1, specifying the data sets that are investigated for BSM signatures in this work.}
    \label{tab:datasets}
\end{table}}%

%%%%%%%%%%%%%%%%%%%%%%%%%%%%%%%%%%%%%%%%%%%%%%%%%%%%%%%%%%%%%%%%%%%%%%%%%%%%%%%%

\subsection{Standard model expectation, likelihood function and systematic uncertainties}
The following investigation relies on a similar analysis chain as the CE$\nu$NS investigation in Ref.~\cite{CONUS:2020skt}.
In this way, we are able to determine realistic bounds on the individual model parameters, while including all relevant experimental uncertainties.
Here we briefly introduce the SM expectations, the performed likelihood procedure and give an overview of the included systematic uncertainties.

The main ingredient of our analysis is a binned likelihood ratio test, cf.~Refs.~\cite{Wilks:1938dza,Cowan:2010js,Lista:2016chp}.
We fix the individual BSM parameters and compare their likelihood value to the one of the null hypothesis, which includes the SM signal of neutrino-nucleus as well as neutrino-electron scattering.
Hence, CE$\nu$NS and neutrino-electron scattering are either modified through interference with new BSM physics or, in the case they are independent, simply appear as an additional background component in the BSM analysis.
From a simulation of the corresponding test statistic (toy MC) we extract limits on these model parameters at 90\,\% confidence level (C.L.).

The differential cross section of the SM predicted CE$\nu$NS process is given by, cf.~Ref.~\cite{Freedman:1973yd},
\begin{align}\label{eq:cross_section_sm_cenns}
	\frac{d\sigma}{d T_{A}}(T_{A}, E_{\nu})= \frac{G_{F}^{2}}{ \pi} \mathcal{Q}^{2}_{W} m_{A} \left(1-
	\frac{m_{A} T_{A}}{2 E^{2}_{\nu}}\right) F^{2}(T_{A})\, ,
\end{align}
with the nuclear recoil energy $T_{A}$, Fermi's constant $G_{F}$, the nuclear mass $m_{A}$ and the neutrino energy $E_{\nu}$.
We use the nuclear charge\footnote{Sometimes, the weak nuclear charge is defined as $\mathcal{Q}_{W} =(1-4\sin^{2}\theta_{W})Z - N$ such that the prefactor of Eq.~\eqref{eq:cross_section_sm_cenns} includes an additional factor of $\frac{1}{4}$.}
\begin{align}\label{eq:nuclear_charge_weak}
	\mathcal{Q}_{W}= g^{p}_{V} Z + g^{n}_{V} N 
	= \left(\frac{1}{2}-2\sin^{2}\theta_{W}\right)Z - \frac{1}{2} N\, ,
\end{align}
with the Weinberg angle $\theta_W$, the number of protons $Z$ and the number of neutrons $N$ in the target nucleus, respectively.
Further, the nuclear form factor $F(T_{A})$ describes the degree of deviation from scattering off a point-like object.
It is approximated with unity for the rest of this analysis which is justified by the small momentum transfer of reactor antineutrinos.
Thus, at a reactor-site the interaction of antineutrinos with the target nuclei can be seen as a process in the fully coherent regime. 
At higher energies, i.e.\ at $\pi$DAR sources, the loss of coherent enhancement is usually described via the form factor parameterization by Helm~\cite{Helm:1956zz} or by Klein and Nystrand~\cite{Klein:1999qj}.
However, the decrease in cross section is small, i.e.\ a factor of $\sim1.4$ for the \textsc{Coherent} experiment~\cite{COHERENT:2018gft}, and introduces only minor uncertainties of $\lesssim 5\%$~\cite{COHERENT:2017ipa,AristizabalSierra:2019zmy}.

Though the (anti)neutrino-electron scattering process $\bar{\nu}_{e} e^{-}\rightarrow \bar{\nu}_{e} e^{-}$ contributes only as a small background to the CE$\nu$NS ROI, it is relevant for our analysis of the light mediator electron channels at higher energies.
The corresponding SM cross section is found to be, cf.~Ref.~\cite{Giunti:2007ry},
\begin{align}\label{eq:cross_section_sm_nu_e}
    \frac{d\sigma}{dT_{e}} (T_{e}, E_{\nu}) = \frac{G_{F}^{2} m_{e}}{2\pi} \Big[ \left(g_{V} + g_{A}\right)^{2} + \left(g_{V} - g_{A}\right)^{2}\left(1-\frac{T_{e}}{E_{\nu}} \right) + \left(g_{A}^{2} - g_{V}^{2}\right)\frac{m_{e} T_{e}}{E_{\nu}^{2}} \Big]\, .
\end{align}
Herein, $T_{e}$ stands for the electron recoil, and $g_{V}=\frac{1}{2} + 2\sin^{2}\theta_{W}$ and $g_{A}=-\frac{1}{2}$ for the effective vector and axial-vector couplings, respectively.\footnote{Generally, the vector and axial-vector couplings to the Z boson are defined as $g_V^{f} = I_{3}^{f} - 2 q^{f} \sin^{2}\theta_{W}$ and $g_{A}=I_{3}^{f}$, respectively. For example, in the case of a muon one obtains $g_{V}^{\mu} = -\frac{1}{2} + 2\sin^{2} \theta_{W}$ and $g_{A}^{\mu} = -\frac{1}{2}$ which reflects a pure neutral current interaction. In case of an electron, there is an additional W boson exchange that enhances the couplings, i.e.\ $g_{V,A}\rightarrow g_{V,A} + 1$. For antineutrinos, the charged current is mediated via a s-channel diagram (instead of a t-channel), which further leads to $g_{A}\rightarrow -g_{A}$.} 
In the case of neutrino-electron scattering, atomic binding effects for recoil energies comparable to atomic binding energies have to be taken into account. 
We follow the procedure proposed in Ref.~\cite{Mikaelyan:2002nv} and apply electron binding energies of germanium taken from Ref.~\cite{Perkins:1991}.

Both interaction channels exhibit a maximum recoil energy obtained from pure forward scattering,
\begin{align}\label{eq:recoil_energy_max}
        T^{\mathrm{max}}_{x}= \frac{2E_{\nu}^{2}}{m_{x} + 2E_{\nu}}\, \quad \text{for } x=\{e,A\}\, .
\end{align}
Note that electron recoils are, contrary to CE$\nu$NS, not affected by quenching, and, thus, the maximal detectable energy, i.e.\ recoil energy subtracted by the electron's binding energy, lies far above the analyzed ROIs. 
For antineutrinos emitted from a reactor core, i.e.\ $E_{\nu}\sim10\MeV$, we obtain maximal recoil energies of $\sim9.9\MeV$ and $\sim3.0\keV$ for electrons and germanium nuclei, respectively.
As a result, SM neutrino-electron scattering features a flat contribution in our ROI whereas the CE$\nu$NS signal rises towards lower energies with a shift in energy according to the underlying quenching factor.

Both cross sections have to be convolved with the reactor antineutrino spectrum $\frac{d N}{d E_{\nu}}$, such that the final number of events is given by
\begin{align}\label{eq:number_of_events}
	N^{\rm SM}_{x} = t \cdot \Phi^{*}\cdot N_{x}^{\mathrm{Ge}} 
	\sum_{i}^{N_{\mathrm{bins}}} \int_{T_{i}-0.5 \Delta T}^{T_{i} + 0.5 \Delta T} dT
	 \int_{E_{\mathrm{min}}}^{E_{\mathrm{max}}} dE_{\nu}\frac{dN}{dE_{\nu}} (E_{\nu}) 
	 \left( \frac{d\sigma}{dT}\right)_{x} \! (T,E_{\nu})
\end{align}
with the experimental lifetime $t$, $N_{x}^{\mathrm{Ge}}$ for $x=\{e, A\}$ as the number of target electrons and nuclei respectively, $N_{\mathrm{bins}}$ the number of spectral bins and $T_{i}$ the energy at the bin center with the bin width $\Delta T$.
The `reduced' reactor flux incorporates all reactor-related quantities and is given by
\begin{align}\label{eq:reduced_flux}
    \Phi^{*}=\frac{P_{\mathrm{th}}}{4\pi d^{2} \bar{E}}\, ,
\end{align}
with the thermal reactor power $P_{\mathrm{th}}$, the detector's distance to the reactor $d$ and the average energy release per fission $\bar{E}$, cf.~Section~\ref{subsec:datasets_experimentalframework}.
The integral over the applied reactor model yields the number of neutrinos emitted per fission and, multiplied with $\Phi^{*}$, gives the expected neutrino flux in units of cm$^{-2}$\,s$^{-1}$ at the experimental site.
Special care has to be taken for the conversion of nuclear recoil energy into detectable signal (ionization energy), which depends on dissipation processes in the chosen detector technology and target material.
To describe this quenching process in germanium, cf.~Section~\ref{subsec:datasets_experimentalframework}, we select three representative $k$-parameter values $k=\{0.12, 0.16, 0.20\}$, i.e.\ spanning the available measured range in the keV$_{ee}$ regime~\cite{Jones:1971ya, Jones:1975zze, Messous:1995dn, Barbeau:2007qi, Barker:2012ek,Scholz:2016qos, Collar:2021fcl}. 
Thereby we make a substantial uncertainty appearing in our analysis explicit.
Finally, the signal expectation has to be convolved with the individual detector response, i.e.\ the energy resolution and the electronic detection efficiency.
For details of the HPGe detectors used within \textsc{Conus}, we refer to our detector publication~\cite{Bonet:2020ntx}.

In our likelihood procedure, \textsc{On} and \textsc{Off} spectra are fitted simultaneously and additional knowledge on parameters is represented by Gaussian pull terms, 
\begin{align}\label{eq:likelihood_full}
    -2\log\, \mathcal{L} = -2\log\, \mathcal{L}_{\rm ON} -2 \log\, \mathcal{L}_{\rm OFF} + 2 \sum_i \frac{(\Theta_{i} - \Theta_i^*)^2}{2 \sigma_{i}^2}\, .
\end{align}
Herein, the parameters $\Theta_i$ of the pull terms have central values $\Theta_i^*$ and uncertainties $\sigma_i$.
The individual detector's noise edge is fitted with an exponential shape parameterized by two free parameters, $\Theta_{\rm thr_{1}}$ and $\Theta_{\rm thr_{2}}$.
For the noise edge description, we refined the exponential function used in Ref.~\cite{CONUS:2020skt} and extended the fit range slightly to lower energy thresholds.
The MC background model, which will be discussed in detail in a separate publication, cf.~Ref.~\cite{CONUSbackground}, represents the physical background components and appears in the likelihood together with a factor $\Theta_{b_{0}}$ that allows for an overall rescaling as well as two additional uncertainties $\Theta_{b_{1,2}}$ allowing for small variations in the shape of the background model. 
These additional degrees of freedom are necessary to incorporate the uncertainties on the production rates of cosmogenic induced isotopes as well as on detector surface effects, i.e.\ from the thickness of the passivation layer. 
The latter especially influences the spectral shape of the background contributions resulting from decays of contaminants on the diode surface such as $^{210}$Pb. 
The corresponding uncertainties do not exceed 5\% and the energy spectrum of the background model is allowed to vary within this range via a second order polynomial distorsion.
Overall, pull terms are assigned to each detector's active volume, its electronic detection efficiency $c_{\rm eff}$, its energy scale calibration uncertainty $\Delta E$ and the reduced flux $\Phi^{*}$.
The uncertainty of the reduced neutrino flux $\Delta \Phi^{*}$ is found to be $\sim \! 3\,\%$, depending on the detector and run, and is dominated by the uncertainty on the reactor thermal power ($\Delta P = 2.3\,\%$)~\cite{Hakenmuller:2019ecb}, the energy released per fission and isotope (cf.~Ref.~\cite{Ma:2012bm}), as well as the detector's distance to the reactor core ($17.1 \pm 0.1$)\,m and correlations among fission fractions (cf.~Ref.~\cite{An:2016srz}).
Summarizing the parameters related to the reactor model as $\Theta_{\rm reactor}$ and the ones related to the detector as $\Theta_{\rm det}$, we can write schematically:
\begin{equation}\label{eq:likelihood_parts}
\begin{aligned}
-2\log\, &\mathcal{L}_{\rm ON}(\Theta_{\rm b_{0,1,2}}, \Theta_{\rm thr_{1,2}}, \Theta_{\rm reactor}, \Theta_{\rm det}, \Theta_{\Delta E} )\, ,\\
-2\log\, &\mathcal{L}_{\rm OFF}(\Theta_{\rm b_{0,1,2}}, \Theta_{\rm thr_{1,2}}, \Theta_{\rm det}, \Theta_{\Delta E})\, .
\end{aligned}
\end{equation}
In Table~\ref{tab:likelihood_uncertainty}, we provide an overview of the uncertainties that enter our likelihood procedure and their approximate size.
{\renewcommand{\arraystretch}{1.5}%
\setlength{\tabcolsep}{0.45cm}%
\begin{table}[t]
    \centering
    \begin{tabularx}{\textwidth}{ 
  >{\centering\arraybackslash}X 
  | >{\centering\arraybackslash}X }
      Quantity & Uncertainty or related parameter \\
      \hline
      background MC & $\Theta_{b_{0}}$ (free), $\Theta_{b_{1}, b_{2}}$ ($\leq 5\,\%$, uncertainty from background model) \\
      noise threshold &  $\Theta_{\rm thr_{1}}$, $\Theta_{\rm thr_{2}}$ (free, uncertainty \hspace{4cm} calculated via toy MC) \\
      reduced neutrino flux $\Delta \Phi^{*}$ & $\sim \! 3\,\%$ \\
      neutrino spectrum & subdominant uncertainty \hspace{4cm}(compared to quenching)\\
      reactor \textsc{On} and \textsc{Off} duration & negligible uncertainty \\
      active mass & $< 1 \,\%$ \\
      electronic detection efficiency $c_{\mathrm{eff}}$ & $\leq 5 \,\%$ \\
      energy calibration uncertainty $\Delta E$ & $15 \eV_{ee}$ \\
      quenching & k (explicitly included) \\
    \end{tabularx}
    \caption{Overview of the quantities entering the likelihood and their corresponding uncertainties. For details and further information see main text.}
    \label{tab:likelihood_uncertainty}
\end{table}}%
Note that the quenching factor is not quoted with an uncertainty as it is the overall dominating systematics and thus is explicitly taken into account by deriving the limits for different $k$-values.

\begin{figure}
	\centering
	\includegraphics[width=0.80\textwidth]{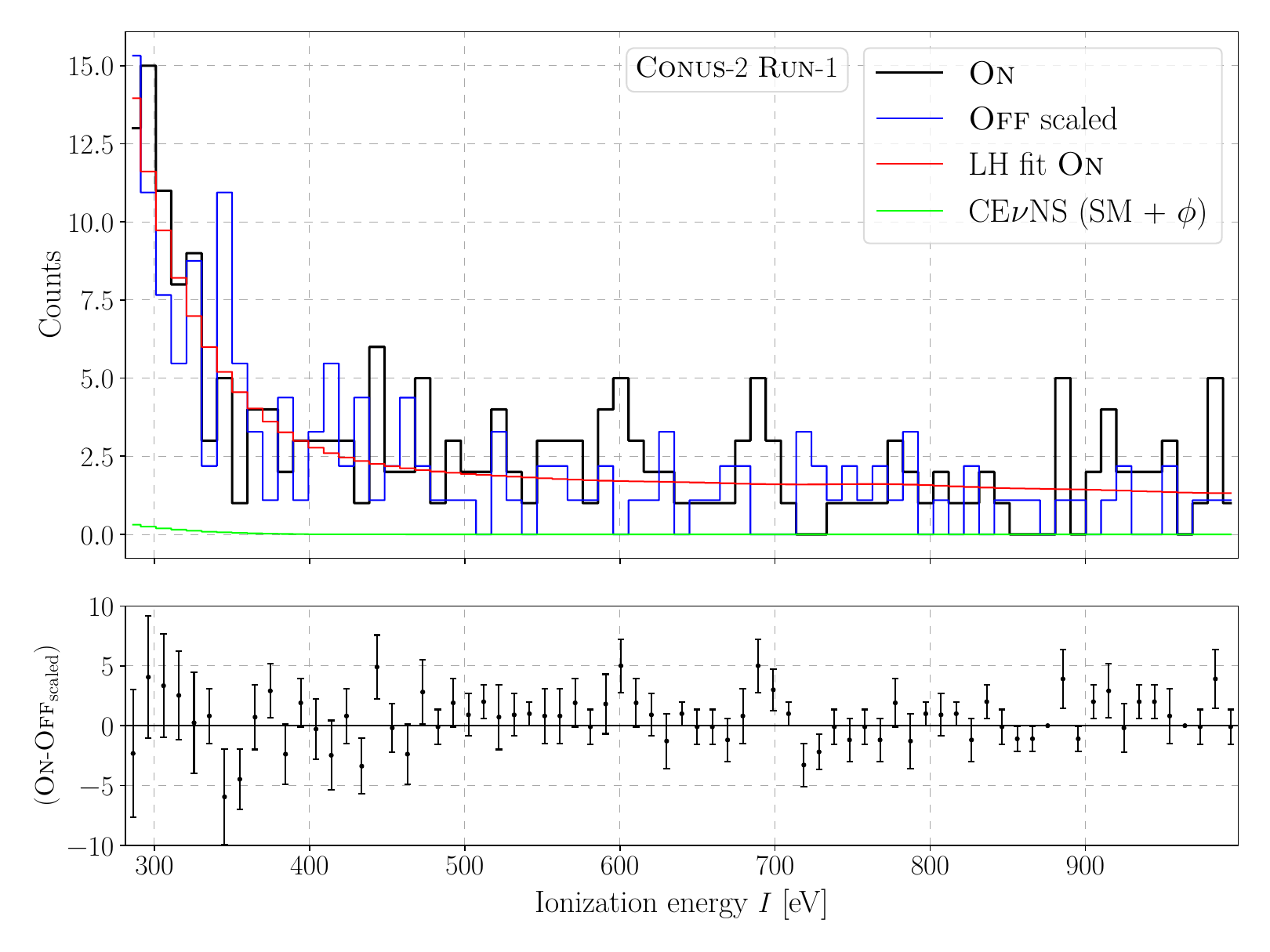}
	\vfill
	\hspace{0.4cm}\includegraphics[width=0.80\textwidth]{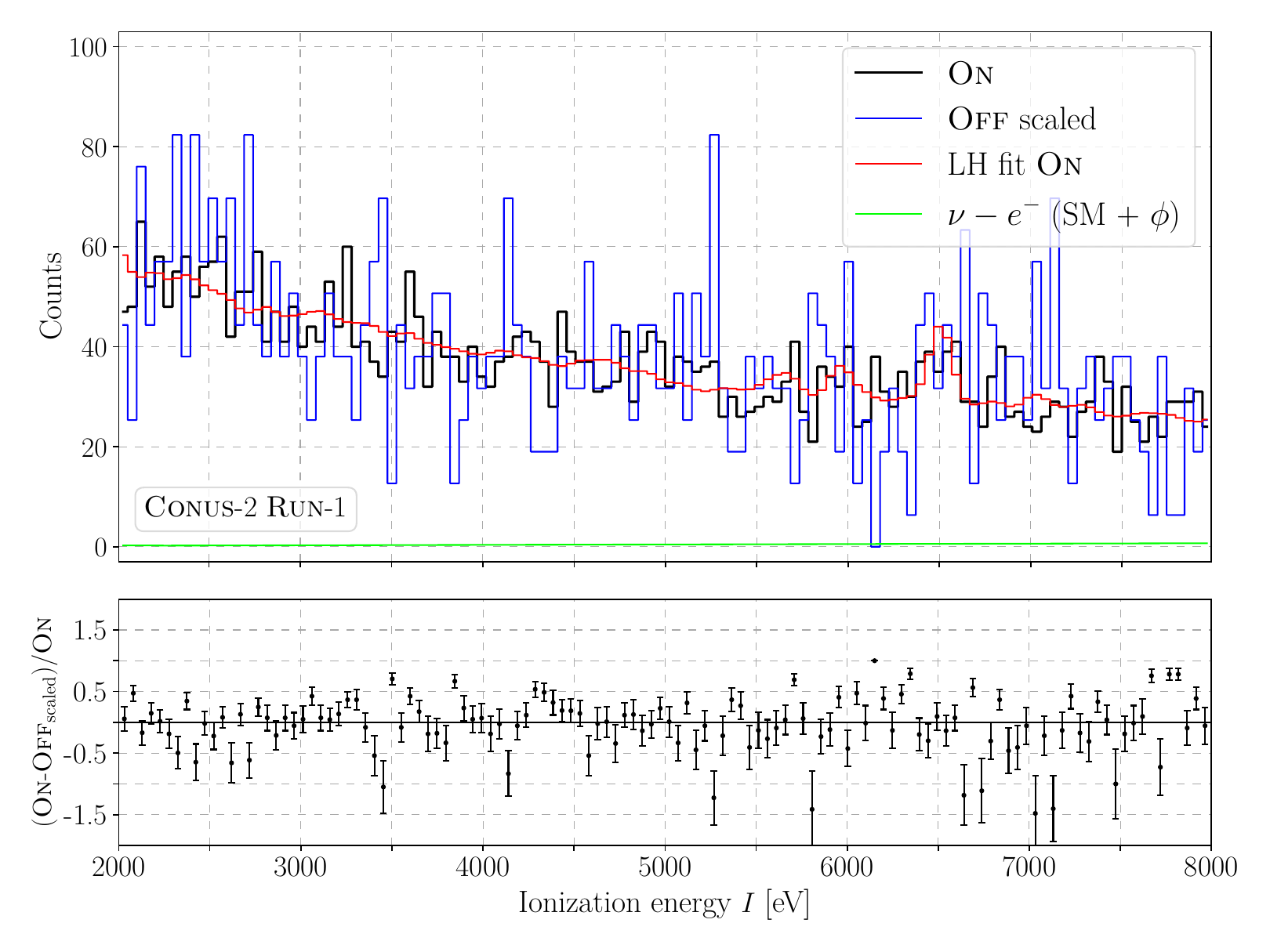}
	\caption{
	Exemplary fits to experimental data in the case of a simplified light scalar mediator, cf.~Section~\ref{sec:light_scalar}.
	A combined fit to all data sets of Table~\ref{tab:datasets} is performed and collected reactor \textsc{On} data (black), the scaled reactor \textsc{Off} data (blue) as well as the obtained likelihood fit (red) are illustrated for detector C2 and a quenching parameter of $k = 0.16$, assuming free coupling and mediator mass of the underlying BSM model. 
	The received signal events (SM + BSM contribution) are indicated in green.
	Top: Fit of the modified CE$\nu$NS signal in the ROI below $1\keV_{ee}$. 
	To illustrate the agreement between the collected reactor \textsc{On} and reactor \textsc{Off} periods, we show the corresponding residuals in total events beneath. 
	Bottom: Fit of modified neutrino-electron scattering in the ROI between $2$ and $8\keV_{ee}$. 
	To quantify the agreement of reactor \textsc{Off} data with the collected \textsc{On} data, residuals are given again (here normalized to the collected \textsc{On} data).
	}
	\label{fig:example_fits}
\end{figure}

The signal hypotheses, which the likelihood compares to the experimental data, are defined by the BSM models described in Section~\ref{sec:bsm_study}.
They are implemented through their corresponding cross sections.
An exemplary (combined) fit to the collected data is illustrated in Figure~\ref{fig:example_fits} for detector C2 and quenching parameter $k=0.16$ in the case of a light scalar mediator, cf.~Section~\ref{sec:light_scalar}.
Contributions to CE$\nu$NS are tested for energies below $1\keV_{ee}$, while the ones to elastic neutrino-electron scattering are examined within an energy range between $2$ and $8\keV_{ee}$.
Further data with their corresponding background models can be found in Refs.~\cite{CONUS:2020skt,CONUSbackground,CONUS:2022qbb}. 
For the minimization of the likelihood we use the iminuit package~\cite{iminuit2020,minuit1975}, while the whole analysis is set up within the SciPy framework~\cite{ScientificComputing2007,ScientificComputing2011,Scipy2020,Matplotlib2007,Ipython2007,Numpy2020,Pandas2010,JupyterLab}. 
The extensive cluster computations are done with the help of the software package MPI for Python~\cite{MPI4Py2005, MPI4Py2008}.

%%%%%%%%%%%%%%%%%%%%%%%%%%%%%%%%%%%%%%%%%%%%%%%%%%%%%%%%%%%%%%%%%%%%%%%%%%%%%%%%

\section{Constraints on beyond the standard model neutrino physics}
\label{sec:bsm_study}

After introducing the experimental characteristics and details of the analysis method, we investigate the \textsc{Conus} \textsc{Run}-1 data set with respect to BSM signatures and compare our results to limits obtained from other CE$\nu$NS experiments.
In particular, we deduce constraints for tensor and vector NSIs as well as simplified light vector and scalar mediators. 
For the latter cases, we can additionally analyze the electron channels of these models with an extended data set at energies between 2 and $8\keV_{ee}$.

%%%%%%%%%%%%%%%%%%%%%%%%%%%%%%%%%%%%%%%%%%%%%%%%%%%%%%%%%%%%%%%%%%%%%%%%%%%%%%%%

%%%%%%%%%%%%%%%%%%%%%%%%%%%%%%%%%%%%%%%%%%%%%%%%%%%%%%%%%%%%%%%%%%%%%%%%%%%%%%%%

%%%%%%%%%%%%%%%%%%%%%%%%%%%%%%%%%%%%%%%%%%%%%%%%%%%%%%%%%%%%%%%%%%%%%%%%%%%%%%%%

\subsection{Non-standard interactions}

%%%%%%%%%%%%%%%%%%%%%%%%%%%%%%%%%%%%%%%%%%%%%%%%%%%%%%%%%%%%%%%%%%%%%%%%%%%%%%%%

A rather model-independent probe of various BSM neutrino physics scenarios are so-called NSIs in the neutrino-quark sector, which are an extension of the neutral current with effective four-fermion operators, generally assuming new mediators that are much heavier than the SM gauge bosons~\cite{Dev:2019anc}.
Since the heavy mediators are conventionally integrated out, the new couplings are defined in terms of Fermi's constant $G_{F}$ analogously to weak interactions at low energy. 
In general, these new couplings can be flavor-preserving $\epsilon_{\alpha \alpha}$ and/or flavor-violating $\epsilon_{\alpha \beta}$ with $\alpha\neq\beta$ and $\alpha,\beta = \{ e, \mu, \tau\}$ being the lepton flavor indices.
Searches of these new neutrino interactions are relevant since they may affect neutrino oscillations~\cite{Farzan:2017xzy} and even other physics branches like cosmology~\cite{Du:2021idh} or astrophysics~\cite{Amanik:2006ad,Stapleford:2016jgz}.
NSIs in their original definition can be studied since they enter the SM CE$\nu$NS cross section via a modified or an additional nuclear charge~\cite{Barranco:2005yy,Barranco:2007tz,Lindner:2016wff}.
More recently, they have been investigated on more general grounds, i.e.\ in the context of so-called general neutrino interactions (GNIs)~\cite{AristizabalSierra:2018eqm,Bischer:2019ttk}.
As \textsc{Conus} operates in the fully coherent regime, the subtleties that can arise for the form factor in BSM models, cf.~Ref.~\cite{Hoferichter:2020osn}, are not of relevance to our analysis here.

%%%%%%%%%%%%%%%%%%%%%%%%%%%%%%%%%%%%%%%%%%%%%%%%%%%%%%%%%%%%%%%%%%%%%%%%%%%%%%%%

\subsubsection{Tensor-type interaction}

%%%%%%%%%%%%%%%%%%%%%%%%%%%%%%%%%%%%%%%%%%%%%%%%%%%%%%%%%%%%%%%%%%%%%%%%%%%%%%%%

\begin{figure}
	\centering
	\includegraphics[width=0.80\textwidth]{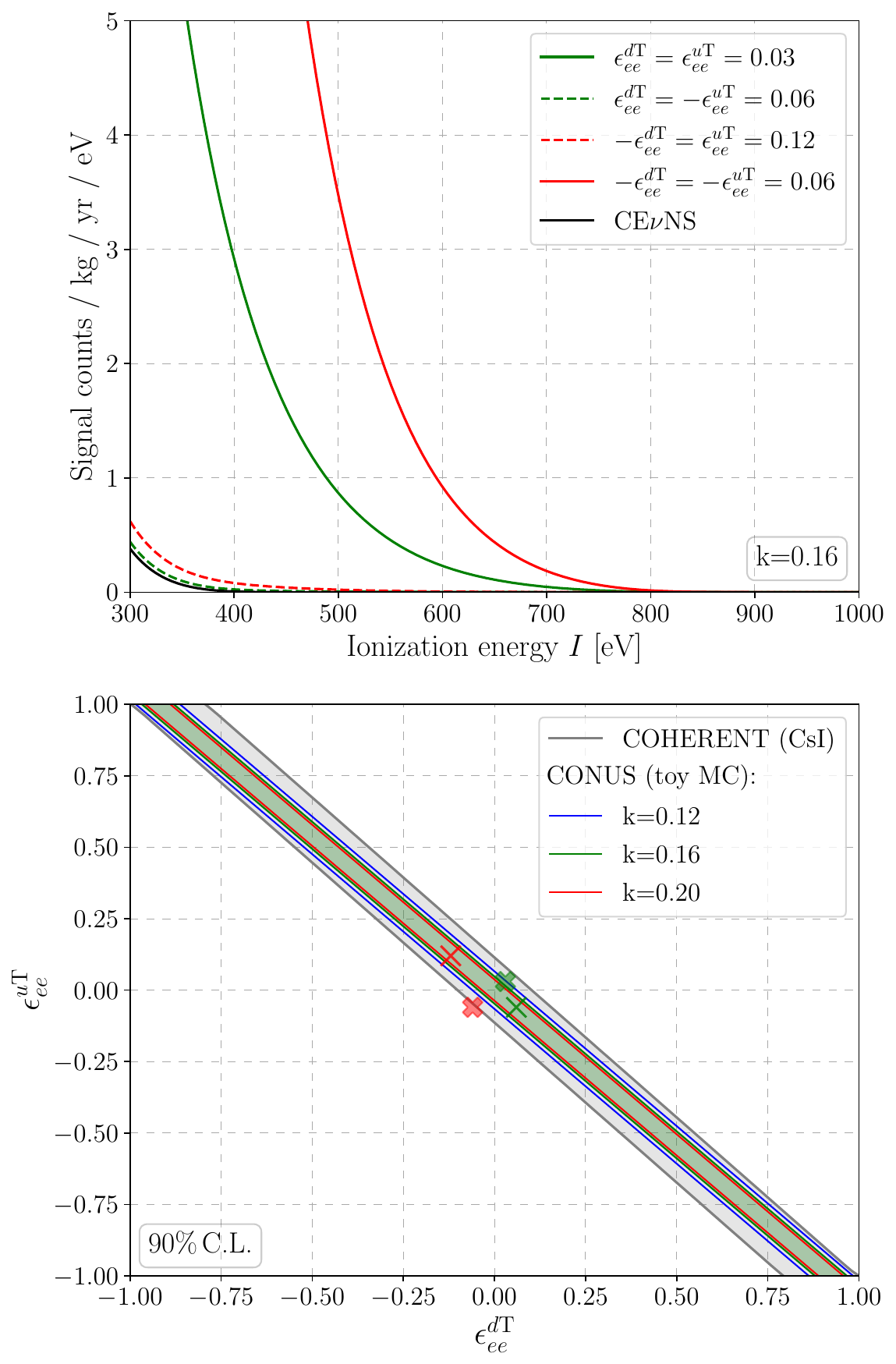}
	\caption{Top: Expected tensor NSI signals of detector C1 for a quenching parameter of $k = 0.16$ and different coupling values from all quadrants in comparison to the standard CE$\nu$NS signal.
	Due to a different chiral structure, additional tensor NSIs can only enhance the expected signal. 
	Bottom: Allowed regions~(at 90\%\,C.L.) of tensor NSI couplings $\epsilon^{u \mathrm{T}}_{ee}$ and $\epsilon^{d\mathrm{T}}_{ee}$ deduced from the \textsc{Run}-1 \textsc{Conus} data set. 
	The exemplary points of the upper plot are marked with crosses, where bold crosses indicate couplings that are (almost) excluded, i.e.\ the solid lines from above.
	Normal crosses refer to coupling combinations that cannot be excluded with the current data set, i.e.\ the dashed lines.
	In addition, constraints~(90\%\,C.L.) obtained from \textsc{Coherent} data are plotted for comparison, cf.~Ref.~\cite{Kosmas:2017tsq}.}
	\label{fig:tensor_nsi_bounds}
\end{figure}

%%%%%%%%%%%%%%%%%%%%%%%%%%%%%%%%%%%%%%%%%%%%%%%%%%%%%%%%%%%%%%%%%%%%%%%%%%%%%%%%

Non-standard neutrino-quark interactions of tensor-type can arise in generalizations of the conventional vector NSI approach~\cite{Barranco:2011wx} and naturally occur in the context of GNIs~\cite{AristizabalSierra:2018eqm,Bischer:2019ttk}.
Furthermore, they might also be associated with electromagnetic properties of neutrinos~\cite{Healey:2013vka,Papoulias:2015iga}.
Here, we assume the existence of new tensor-type interactions between neutrinos and quarks which are induced by an operator of the form
\begin{align}\label{eq:operator_tensor_nsi}
\mathcal{O}^{q \mathrm{T}}_{\alpha\beta} = \left( \bar{\nu}_{\alpha} \sigma^{\mu \nu} \nu_{\beta}\right)\left( \bar{q}\sigma_{\mu \nu} q \right) + \text{h.c.}\, ,
\end{align}
with $q$ denoting the first generation of quarks $q=\{u,d\}$ and $\alpha,\beta = \{ e, \mu, \tau\}$ being the lepton flavor indices.
Due to a different chiral structure, there is no possibility of destructive interference with the SM channel.
The corresponding couplings to quarks can be combined into a new nuclear charge in equivalence to the SM weak charge appearing in the CE$\nu$NS cross section of Eq.~\eqref{eq:cross_section_sm_cenns}.
Thus, in our case we have 
\begin{align}\label{eq:nuclear_charge_tensor_nsi}
\mathcal{Q}_{\rm NSI}^{\mathrm{T}} = \left( 2\epsilon^{u \mathrm{T}}_{\alpha\beta} + \epsilon^{d \mathrm{T}}_{\alpha\beta} \right) Z +  \left( \epsilon^{u \mathrm{T}}_{\alpha\beta} + 2\epsilon^{d \mathrm{T}}_{\alpha\beta} \right) N\, ,
\end{align}
with the lepton flavor indices $\alpha,\beta$ as well as $Z$ and $N$ representing the respective number of protons and neutrons in the target nucleus.
Note that in contrast to the SM case, cf.~Eq.~\eqref{eq:cross_section_sm_cenns} and Eq.~\eqref{eq:nuclear_charge_weak}, here, as well as in the other BSM models, the proton number does not get weighted with a small prefactor.
Thus, the cross section does not necessarily scale with the characteristic dependence on the squared neutron number.
Although flavor-changing tensor-type interactions can in principle appear and are for example tested at $\pi$DAR sources~\cite{Kosmas:2017tsq}, at reactor site we are only able to probe couplings related to the electron flavor.
Therefore, in this analysis, we focus on flavor-diagonal couplings, i.e.\ $\epsilon^{u \mathrm{T}}_{ee}$ and $\epsilon^{d \mathrm{T}}_{ee}$.

The new tensor-type interaction simply adds to the conventional CE$\nu$NS cross section, resulting in, cf.~Ref.~\cite{Papoulias:2015iga},
\begin{align}\label{eq:cross_section_tensor_nsi_cenns}
	\left(\frac{d\sigma}{dT_{A}}\right) = \left(\frac{d\sigma}{dT_{A}}\right)_{\rm CE\nu NS} + \frac{4 G^{2}_{F}}{\pi} {\mathcal{Q}^{ \mathrm{T}}_{\rm NSI}}^{2} \, m_N
	\left(1- \frac{m_{A} T_{A}}{4 E^{2}_{\nu}}\right)\, .
\end{align}
Note the different kinematic factors between the CE$\nu$NS cross section in Eq.~\eqref{eq:cross_section_sm_cenns} and Eq.~\eqref{eq:cross_section_tensor_nsi_cenns} which allow the tensor NSI signal to extend to higher energies.
The upper plot of Figure~\ref{fig:tensor_nsi_bounds} illustrates the modified signal expectation in detector C1 due to additional tensor NSIs in comparison to the SM case.
It shows when up- and down-quark couplings have different signs, the amplitude of the BSM signal is significantly smaller than in the case of same signs.

The obtained limits at 90\% C.L.\ for tensor NSIs from the analysis of the \textsc{Conus} \textsc{Run}-1 data are shown in the lower plot of Figure~\ref{fig:tensor_nsi_bounds}, where they are compared with similar bounds deduced from CsI(Na) data of the \textsc{Coherent} experiment.\footnote{For the extraction of limits shown throughout this paper, we used the tool WebPlotDigitizer~\cite{Rohatgi2020}.}
For illustrative purposes, the parameter points of the example BSM signal rates, shown in the upper plot of Figure~\ref{fig:tensor_nsi_bounds}, are marked with crosses.
Although \textsc{Conus} has not observed a CE$\nu$NS signal yet, we place competitive bounds on the tensor NSI couplings $\epsilon^{u \mathrm{T}}_{ee}$ and $\epsilon^{d \mathrm{T}}_{ee}$.\footnote{
Note that the indices here, referring to the electron (anti)neutrinos involved in the new scattering process, are not to be confused with the indices of eV$_{ee}$, referring to the ionization energy.}
This is due to the signal's higher extent (compared to SM CE$\nu$NS) and the low background levels obtained below $1\keV_{ee}$. 
Here, the quenching factor's impact is of minor importance since, for the values considered, the tensor NSI signal lies way above the \textsc{Conus} energy threshold allowing for bounds that are mainly dominated by the experimental conditions like background and exposure. 
Figure~\ref{fig:tensor_nsi_bounds} furthermore illustrates how the degeneracy between the two NSI couplings, $\epsilon^{u \mathrm{T}}_{ee}$ and $\epsilon^{d \mathrm{T}}_{ee}$, can be broken.
The different slopes of the limit bands that are visible for \textsc{Conus} and \textsc{Coherent} are due to the different detector isotopes used in the experiments.
In general, they allow for breaking the degeneracy of the couplings.
However, with data obtained so far the difference between the detector materials CsI and Ge (in terms of $N$ and $Z$) is not sufficient to have a substantial impact on the combined allowed regions.

Since NSIs are by definition induced by a new heavy mediator that has been integrated out, we can translate the bounds we found for the tensor NSIs into a scale at which this effective description is expected to break down.
This scale, where new physics gets probed, is given by $\Lambda \approx g_{x}/g \cdot M_{W}/\sqrt{\epsilon} \sim M_{W}/\sqrt{\epsilon}$, cf.~Ref.~\cite{Lindner:2016wff}, and, in the case of our determined limits, turns out to be higher than $\sim360\GeV$.
Hence, with increasing sensitivity low energy experiments like \textsc{Conus} might probe physics at energy scales comparable to the \textsc{Lhc} (TeV scale).

%%%%%%%%%%%%%%%%%%%%%%%%%%%%%%%%%%%%%%%%%%%%%%%%%%%%%%%%%%%%%%%%%%%%%%%%%%%%%%%%

\subsubsection{Vector-type interaction}

%%%%%%%%%%%%%%%%%%%%%%%%%%%%%%%%%%%%%%%%%%%%%%%%%%%%%%%%%%%%%%%%%%%%%%%%%%%%%%%%

\begin{figure}
	\centering
	\includegraphics[width=0.8\textwidth]{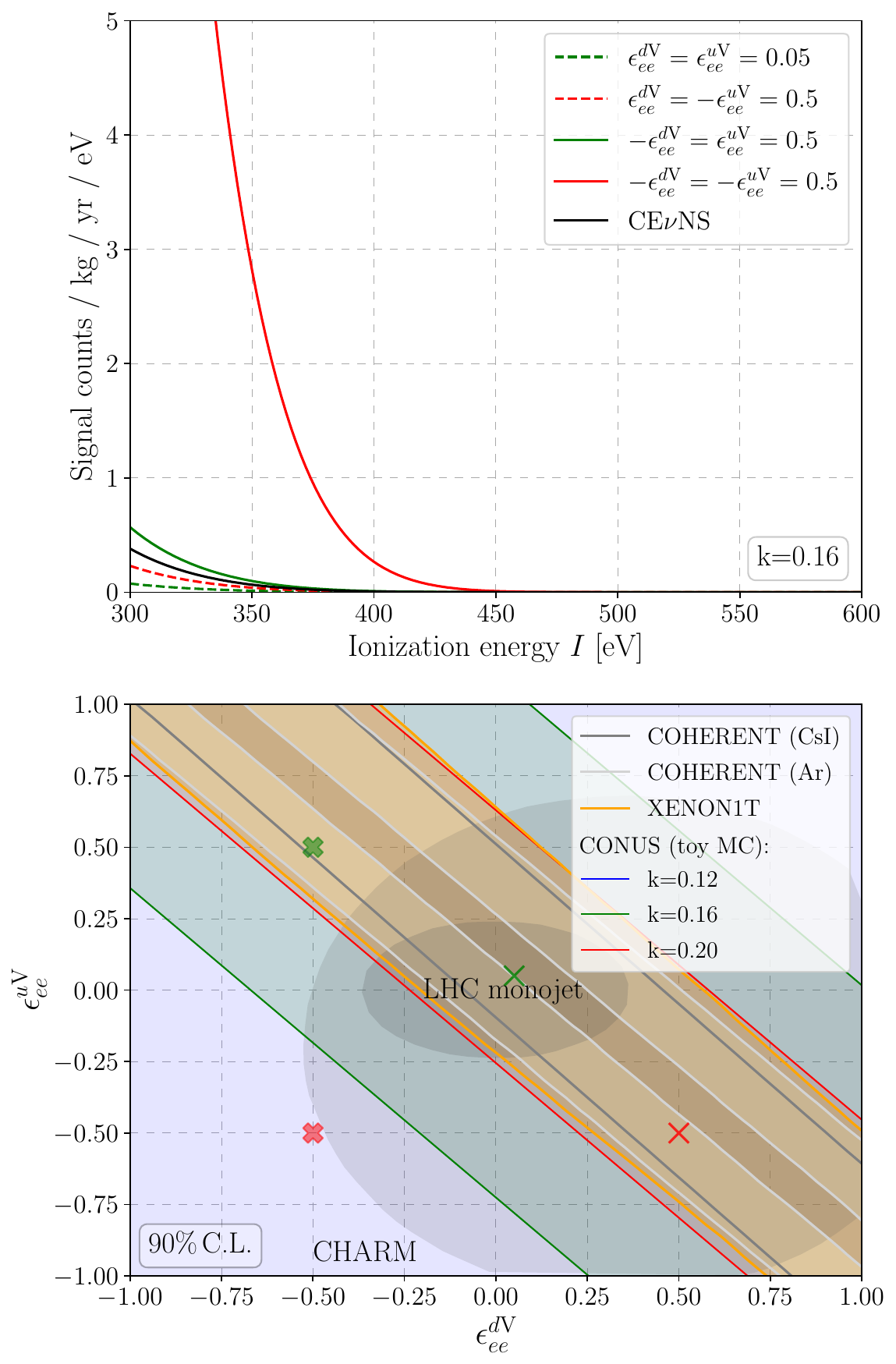}
	\caption{Top: Expected vector NSI signals in detector C1, assuming a quenching parameter of $k = 0.16$ and different coupling values from all quadrants in comparison to the standard CE$\nu$NS signal.
	Note that, depending on the explicit couplings, destructive interference between the vector NSIs and the SM signals is possible and the expected number of events can be reduced (with respect to the pure SM case).
	Bottom: Allowed regions~(at 90\%\,C.L.) of vector NSI couplings $\epsilon^{u \mathrm{V}}_{ee}$ and $\epsilon^{d \mathrm{V}}_{ee}$ deduced from the \textsc{Run}-1 \textsc{Conus} data set. 
	As in Figure~\ref{fig:tensor_nsi_bounds}, the example points of the upper plot are marked with crosses, where bold crosses indicate signals stronger than the SM expectation and normal crosses point to the parameter space of destructive interference between the SM and BSM channels.
	For comparison, constraints (90\%\,C.L.) obtained from \textsc{Coherent} (CsI~\cite{Papoulias:2019txv} and Ar~\cite{COHERENT:2020iec}) data and the Xenon1T experiment~\cite{XENON:2020gfr} are shown.
	Further existing limits, e.g.\ from \textsc{Charm}~(90\%\,C.L.)~\cite{Dorenbosch:1986tb} and \textsc{Lhc} monojet searches~(95\%\,C.L.)~\cite{Friedland:2011za} are indicated with grey elliptic regions.
	}
	\label{fig:vector_nsi_bounds}
\end{figure}

%%%%%%%%%%%%%%%%%%%%%%%%%%%%%%%%%%%%%%%%%%%%%%%%%%%%%%%%%%%%%%%%%%%%%%%%%%%%%%%%

Using the same notation as for the tensor-type NSIs, the vector-type NSIs represent a four-fermion interaction described by the operator
\begin{align}\label{eq:operator_vector_nsi}
    \mathcal{O}^{q \mathrm{V}}_{\rm NSI}=\left(\bar{\nu}_{\alpha} \gamma^{\mu} L \nu_{\beta}
 \right)\left(\bar{q} \gamma_{\mu}Pq  \right) + \text{h.c.}\, ,
\end{align}
with left- and right-handed projection operators $P = \{L,R\}$.
Since this new vector-type interaction exhibits a structure similar to the conventional SM CE$\nu$NS, the related couplings to quarks can be directly absorbed in the weak charge, cf.~Eq.~\eqref{eq:cross_section_sm_cenns}: $\mathcal{Q}_{W} \rightarrow \mathcal{Q}_{\rm NSI}^{\mathrm{V}}$.
Furthermore, the operator in Eq.~\eqref{eq:operator_vector_nsi} can trigger a flavor change among the involved neutrinos and, thus, neutrino-nucleus scattering might become flavor-dependent.
In its most general version, the modified weak charge now reads, cf.~Ref.~\cite{Barranco:2005yy},
\begin{equation}
\begin{aligned}\label{eq:nuclear_charge_vector_nsi}
	\mathcal{Q}_{\rm NSI}^{\mathrm{V}} =& \left( g_{V}^{p} + 2 \epsilon^{u\mathrm{V}}_{\alpha \alpha} 
	+ \epsilon^{d\mathrm{V}}_{\alpha \alpha} \right) Z + \left(g_{V}^{n} + \epsilon^{u\mathrm{V}}_{\alpha \alpha} + 2 \epsilon^{d\mathrm{V}}_{\alpha \alpha}\right) N \\
	&+ \sum_{\alpha, \beta} \left[ \left( 2\epsilon^{u\mathrm{V}}_{\alpha\beta} + \epsilon^{d\mathrm{V}}_{\alpha\beta} \right)Z + \left( \epsilon^{u\mathrm{V}}_{\alpha \beta } + 2 \epsilon^{d\mathrm{V}}_{\alpha \beta} \right)N  \right]\, ,
\end{aligned}
\end{equation}
where the first line represents the flavor-preserving interactions (including SM CE$\nu$NS) and the second line the flavor-changing interactions.
As for tensor NSIs, with reactor antineutrinos it is only possible to probe effective couplings of electron-type, i.e.\ $\epsilon^{u \mathrm{V}}_{ee}$ and $\epsilon^{d \mathrm{V}}_{ee}$.
In contrast, with $\pi$-DAR beams it is possible to investigate several types of couplings since they contain muon (anti)neutrinos as well.
Investigations of the \textsc{Coherent} data have already led to bounds on such couplings, either assuming one to be non-vanishing at a time, e.g.~Ref.~\cite{Khan:2019cvi,Papoulias:2019txv}, or in a combined approach with oscillation data that takes into account flavor-changing couplings as well, cf.~Ref.~\cite{Coloma:2019mbs}.

The expectation of potential vector NSI signals within detector C1 are shown in the upper plot of Figure~\ref{fig:vector_nsi_bounds} together with the corresponding SM CE$\nu$NS signal.
Both signals share the same kinematic cut-off and due to the same chiral structure, destructive interference is possible in some regions of the parameter space.
Thus, (CE$\nu$NS + vector NSI) signal rates smaller than the expected CE$\nu$NS rate alone are possible in the context of vector NSIs as indicated by the dashed lines in the upper plot of Figure~\ref{fig:vector_nsi_bounds}.

In contrast to tensor NSIs, the vector NSI case does not benefit from an extent to higher energies.
As a consequence, we cannot hope to obtain equally strong bounds as \textsc{Coherent}. 
This effect is visible in the lower plot of Figure~\ref{fig:vector_nsi_bounds}, which shows the deduced limits on vector NSIs from the \textsc{Conus} \textsc{Run}-1 data set in comparison to the existing limits, i.e.\ from the experiments \textsc{Coherent} and Xenon1T.
It is apparent that the strength of the limits for vector NSIs strongly depends on the quenching factor, which is due to the fact that the quenching factor significantly influences the expected number of events in the ROI.
Comparing the derived \textsc{Conus} limits on vector NSIs for the currently favored quenching value of $k=0.16$ to bounds from other experiments, we find that they are currently subdominant.
Furthermore, resolving the region of destructive interference is beyond the current experimental reach.
However, further experimental improvements that could lead to a future detection of CE$\nu$NS would also significantly improve the sensitivity to vector NSIs and could even allow to probe the parameter region of strong destructive interference.

%%%%%%%%%%%%%%%%%%%%%%%%%%%%%%%%%%%%%%%%%%%%%%%%%%%%%%%%%%%%%%%%%%%%%%%%%%%%%%%%

\subsection{Simplified mediator models}

Another class of models that can be constrained with \textsc{Conus} data are so-called `simplified models' that have been intensively studied, e.g.\ in the dark matter searches at the \textsc{Lhc}~\cite{Abdallah:2015ter,Abercrombie:2015wmb,Boveia:2016mrp}.
Although such kind of models have to be taken with care~\cite{Kahlhoefer:2015bea,Ellis:2017tkh,Morgante:2018tiq}, they experience great popularity since they do not need to be fully specified at high energy. 
Besides dark matter and neutrino physics, this simple framework is applied in various contexts, such as in searches for two Higgs doublet models at the \textsc{Lhc}~\cite{Arcadi:2020gge} or for leptoquark investigations of B-mesons anomalies~\cite{Vignaroli:2019lkg}. 
For neutrino-electron scattering or neutrino-nucleus scattering measurements, such models are interesting since the mediators can have an impact on the recorded recoil spectra, most pronounced for mediator masses that are smaller than the maximal momentum transfer.
Thus, experiments using reactor antineutrinos can, especially in the mediator mass region below $\sim \! 10\MeV$, be even more sensitive than experiments using $\pi$-DAR sources.
In the following, we investigate signatures of new scalar and vector mediators that might scatter off nuclei or electrons by using the \textsc{Conus} \textsc{Run}-1 data sets as defined in Table~\ref{tab:datasets}.

%%%%%%%%%%%%%%%%%%%%%%%%%%%%%%%%%%%%%%%%%%%%%%%%%%%%%%%%%%%%%%%%%%%%%%%%%%%%%%%%

\subsubsection{Light vector bosons}

%%%%%%%%%%%%%%%%%%%%%%%%%%%%%%%%%%%%%%%%%%%%%%%%%%%%%%%%%%%%%%%%%%%%%%%%%%%%%%%%

\begin{figure}
	\centering
	\includegraphics[width=0.80\textwidth]{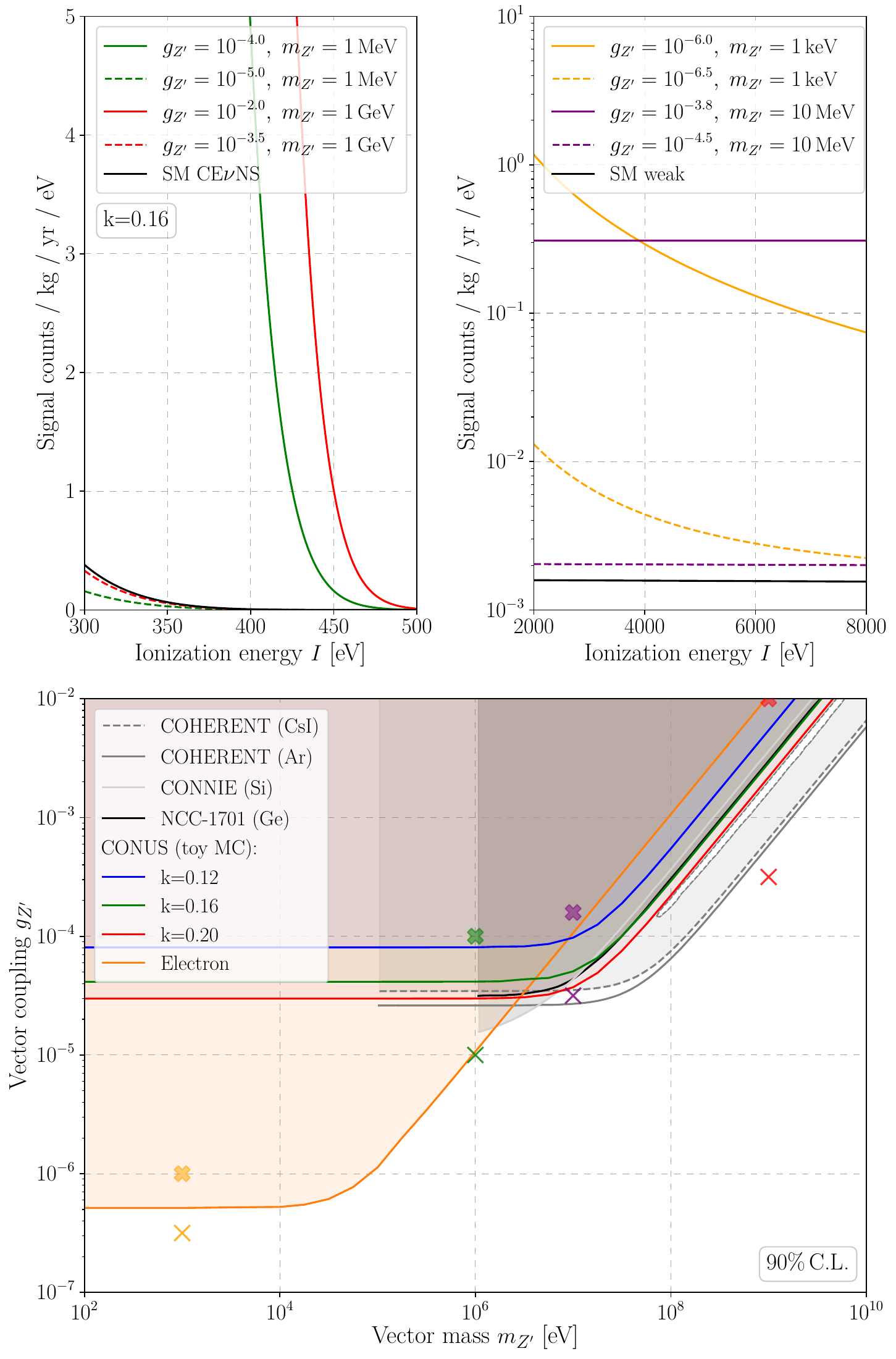}
	\caption{Top: Expected light vector signals of detector C1 in the low energy region below $500 \eV_{ee}$ for a quenching parameter of $k = 0.16$ (left) and in the higher energy region between $2$ and $8 \keV_{ee}$ (right) for different couplings and masses in comparison to the SM signals of CE$\nu$NS and elastic neutrino-electron scattering, respectively.
	Bottom: Limits~(90\%\,C.L.) on the light vector mediator parameters $(m_{Z'}, g_{Z'})$ deduced from CE$\nu$NS and neutrino-electron scattering with the \textsc{Run}-1 \textsc{Conus} data sets. 
	The exemplary parameter points of the upper signal spectra are shown as well.
	Bold crosses indicate parameter points that can already be excluded while regular crosses refer to points that are still allowed.
	For comparison, limits obtained from \textsc{Coherent} (CsI and Ar) data~(90\%\,C.L.)~\cite{Cadeddu:2020nbr}, \textsc{Connie}~(95\%\,C.L.)~\cite{Aguilar-Arevalo:2019zme} as well as \textsc{Ncc-1701}~(95\%\,C.L., quenching according to Ref.~\cite{Collar:2021fcl})~\cite{Colaresi:2021kus} are shown.
	The `island of non-exclusion' in the \textsc{Coherent} limits is due to destructive interference and does not appear in the \textsc{Connie}, \textsc{Conus} and \textsc{Ncc-1701} limits as these experiments have not yet reached the necessary sensitivity.
	}
	\label{fig:light_vector_constraints}
\end{figure}

%%%%%%%%%%%%%%%%%%%%%%%%%%%%%%%%%%%%%%%%%%%%%%%%%%%%%%%%%%%%%%%%%%%%%%%%%%%%%%%%

New $Z$-like vector bosons arise in simple $U(1)$ extensions of the SM and have been studied in various scenarios such as  gauged $B-L$, sequential SM and multiple others, cf.~e.g.~Refs.~\cite{Bauer:2018onh,Arcadi:2017hfi}.
Setting the model-building aside, we can work with an effective Lagrangian including vector-type interactions of neutrinos, quarks and electrons, of the form 
\begin{align}\label{eq:lagrangian_vector_mediator}
	\mathcal{L}_{Z'} = Z'_{\mu} \left( g^{\nu \mathrm{V}}_{Z'} \bar{\nu}_{L} \gamma^{\mu} \nu_{L}
	+  g^{e \mathrm{V}}_{Z'} \bar{e} \gamma^{\mu} e  + g^{q \mathrm{V}}_{Z'} \bar{q} \gamma^{\mu} q \right) 
	+ \frac{1}{2} m^{2}_{Z'} Z'_{\mu}Z'^{\mu}\, ,
\end{align}
with vector-type couplings $g^{x \mathrm{V}}_{Z'}$ ($x = \{\nu,\, e,\, q\}$ and $q = \{u,\, d\}$) and mass of the new vector boson $m_{Z'}$. 
Within this simplified model, we only include interactions of SM neutrinos, i.e.\ left-handed neutrinos and right-handed antineutrinos, and do not take into account characteristic features like kinetic or mass mixing.
In the following, we investigate two reaction channels that arise from Eq.~\eqref{eq:lagrangian_vector_mediator}: neutrino-nucleus as well as neutrino-electron scattering.
In both cases, the light vector boson adds a new reaction channel that can interfere with the SM one, since both share the same final state.
For our investigation, we assume universal couplings, i.e.\ $g_{Z'} \equiv g^{\nu \mathrm{V}}_{Z'}=g^{e \mathrm{V}}_{Z'}=g^{u \mathrm{V}}_{Z'}=g^{d \mathrm{V}}_{Z'}$, allowing us to reduce the parameter space to only two parameters:  $(m_{Z'},\, g_{Z'})$.

The cross section of neutrino-nucleus scattering including a light vector contribution can be expressed as~\cite{Kosmas:2017tsq}
\begin{align}\label{eq:cross_section_light_vector_cenns}
	\left(\frac{d\sigma}{dT_{A}}\right)_{\rm CE\nu NS \, + \, Z'} = \mathcal{G}_{Z'}^{2}(T_{A})\left(\frac{d\sigma}{dT_{A}}\right)_{\rm CE\nu NS}\, , 
\end{align}
with the SM cross section as given in Eq.~\eqref{eq:cross_section_sm_cenns} and the prefactor $\mathcal{G}_{Z'}$ defined as\footnote{In literature, other definitions of $\mathcal{G}_{Z'}$ can be found which differ by a factor of $\frac{1}{2}$. As mentioned before, these differences are due to different definitions of $\mathcal{Q}_{W}$.}
\begin{align}\label{eq:prefactor_light_vector}
    \mathcal{G}_{Z'}(T_{A}) = 1	+ \frac{g^{\nu \mathrm{V}}_{Z'}}{\sqrt{2}G_{F}} \frac{\mathcal{Q}_{Z'}}{\mathcal{Q}_{W}} \frac{1}{2 m_{A} T_{A} + m^{2}_{Z'}}\, .
\end{align}
The nuclear charge associated to the light vector mediator is given by~\cite{Bertuzzo:2017tuf}
\begin{align}\label{eq:nuclear_charge_light_vector}
    \mathcal{Q}_{Z'} = \left(2g^{u \mathrm{V}}_{Z'} + g^{d \mathrm{V}}_{Z'}\right)Z + \left( g^{u \mathrm{V}}_{Z'} + 2 g^{d \mathrm{V}}_{Z'}\right) N \rightarrow 3\, g_{Z'} \left( Z+N \right)\, ,
\end{align}
where the last step is due to our assumption of universal couplings to leptons and quarks.
As a result, the light vector part of Eq.~\eqref{eq:prefactor_light_vector} scales as $g_{Z'}^{2}$, leading to a proportionality of up to $g_{Z'}^{4}$ in the cross section of Eq.~\eqref{eq:cross_section_light_vector_cenns}.
A second effect that becomes visible in Eq.~\eqref{eq:prefactor_light_vector} is the possibility of destructive interference, originating from a negative coupling, which leads to `islands of non-exclusion' in the exclusion plot, cf.~\textsc{Coherent} limits in Figure~\ref{fig:light_vector_constraints}.
In this case the prefactor $\mathcal{G}_{Z'}$ turns from the SM value $1$ into $-1$ due to the $Z'$ contribution, leaving the resulting cross section invariant, cf.~Eq.~\eqref{eq:cross_section_light_vector_cenns}.
However, reactor experiments do not have the sensitivity to observe this effect yet, cf.~Figure~\ref{fig:light_vector_constraints}.

It is worth to mention that there is in principle a connection between the vector mediators discussed here and the previously discussed vector NSIs.
Integrating out the vector mediator allows for a mapping between the $Z'$ couplings and mass and the $\epsilon$-parameters of vector NSIs~\cite{Denton:2018xmq}
\begin{align}
\epsilon^{q \mathrm{V} }_{\alpha \beta} =\frac{\left( g_{Z'}^{\nu \mathrm{V} }\right)_{\alpha\beta} g^{q \mathrm{V}}_{Z'}}{2\sqrt{2}G_{F} M^{2}_{Z'}}\, ,
\end{align}
where the couplings $\left(g_{Z'}^{\nu V}\right)_{\alpha\beta}$ can in general be flavor-dependent.
However, integrating out the mediating particle is only possible when the mediator is significantly heavier than the momentum transfer in the scattering process.
Since this condition is violated for light mediators, we discuss the two models separately.

In addition to neutrino-nucleus scattering, we also look at the influence of a new vector mediator on neutrino-electron scattering.
The corresponding cross section is given by~\cite{Cerdeno:2016sfi}
\begin{align}\label{eq:cross_section_ligth_vector_nu_e}
    \left(\frac{d\sigma}{dT_{e}}\right)_{\nu e + Z'}=\left(\frac{d\sigma}{dT_{e}}\right)_{\nu e} 
    + \frac{\sqrt{2} G_{F} m_{e} g_{V} g_{Z'}^{\nu \mathrm{V}} g_{Z'}^{e \mathrm{V}} }{\pi (2m_{e}T_{e}  + m_{Z'}^{2})} 
    + \frac{m_{e} (g_{Z'}^{\nu \mathrm{V}} g_{Z'}^{e \mathrm{V}})^{2} }{2\pi (2m_{e}T_{e}  + m_{Z'}^{2})^{2}}\, ,
\end{align}
with the electron vector coupling to Z bosons $g_{V} = -\frac{1}{2} + 2 \sin^{2}\theta_{W}$.
By comparing the last term of Eq.~\eqref{eq:cross_section_ligth_vector_nu_e} to Eq.~\eqref{eq:cross_section_light_vector_cenns}, we can see how neutrino-electron scattering can enable us to set stronger limits for small $Z'$ masses.
For $m_{Z'}^2 \ll 2m_{e}T_{e}$, the electron mass $m_e$ in the numerator cancels out and we end up with $4 m_e T_e^2$ in the denominator.
Comparing this to the denominator $4 m_{A} T_{A}^2$ in Eq.~\eqref{eq:cross_section_light_vector_cenns} (together with Eq.~\eqref{eq:cross_section_sm_cenns}), we note that the smaller electron mass enhances our cross section and thus leads to a stronger limit for universal couplings in this region of our parameter space.

Exemplary event spectra for neutrino-nucleus and neutrino-electron scattering for detector C1 are shown in the upper plots of Figure~\ref{fig:light_vector_constraints} for two different masses of the $Z'$ and two different couplings for each mass.
The conventional SM channels are illustrated for comparison.
Especially, note the change in shape for elastic neutrino-electron scatterings of the shown parameter points in the upper right plot of Figure~\ref{fig:light_vector_constraints} which illustrates the different behavior for the denominator in Eq.~\eqref{eq:cross_section_ligth_vector_nu_e} mentioned above. 
In the lower plot of Figure~\ref{fig:light_vector_constraints}, the resulting limits of our analysis are depicted in the ($m_{Z'}$,\,$g_{Z'}$)-plane together with bounds from \textsc{Coherent}~\cite{Dutta:2019eml,Cadeddu:2020nbr,Miranda:2020zji,Kosmas:2017tsq}, \textsc{Connie}~\cite{Aguilar-Arevalo:2019zme} and \textsc{Ncc-1701}~\cite{Colaresi:2021kus}.
For $Z'$ masses above $10 \MeV$, the strongest bounds can be set by $\pi$DAR experiments because of their higher neutrino energies, while for smaller masses reactor experiments can set competitive or stronger bounds.
Furthermore, the limits we can set from neutrino-electron scattering are stronger than the ones from neutrino-nucleus scattering for $m_{Z'} \lesssim 10\MeV$ as explained before.
With the current data set and the most favored quenching value $k=0.16$, the lowest coupling value that can be probed with CE$\nu$NS is $\sim 4 \cdot10^{-5}$.
In the case of elastic neutrino-electron scattering the coupling can be constrained down to $\sim 6\cdot10^{-7}$ for lowest mediator masses.

Besides the bounds from CE$\nu$NS experiments shown in Figure~\ref{fig:light_vector_constraints}, there exists a plethora of bounds on vector mediators from various other types of experiments, especially in the context of a gauged $U(1)_{B-L}$ symmetry.
This includes searches for dielectron resonances at \textsc{Atlas}~\cite{Aaboud:2016cth}, beam dump investigations~\cite{Batell:2009di,Bjorken:2009mm}, bounds from neutrino-electron scattering~\cite{Bilmis:2015lja,Lindner:2018kjo} as well as dark photon searches at BaBar~\cite{Lees:2014xha,Lees:2017lec} and LHCb~\cite{Aaij:2017rft}.
Numerous collections of bounds can be found e.g.\ in Refs.~\cite{Ilten:2018crw,Bauer:2018onh} for general models and Ref.~\cite{Harnik:2012ni} for $B-L$ extensions.
While focusing on the strengths of the limits derived in this work in context of CE$\nu$NS experiments, we mention the broader scope of bounds for the interested reader.

%%%%%%%%%%%%%%%%%%%%%%%%%%%%%%%%%%%%%%%%%%%%%%%%%%%%%%%%%%%%%%%%%%%%%%%%%%%%%%%%

\subsubsection{Light scalar bosons}\label{sec:light_scalar}

%%%%%%%%%%%%%%%%%%%%%%%%%%%%%%%%%%%%%%%%%%%%%%%%%%%%%%%%%%%%%%%%%%%%%%%%%%%%%%%%

\begin{figure}
	\centering
	\includegraphics[width=0.80\textwidth]{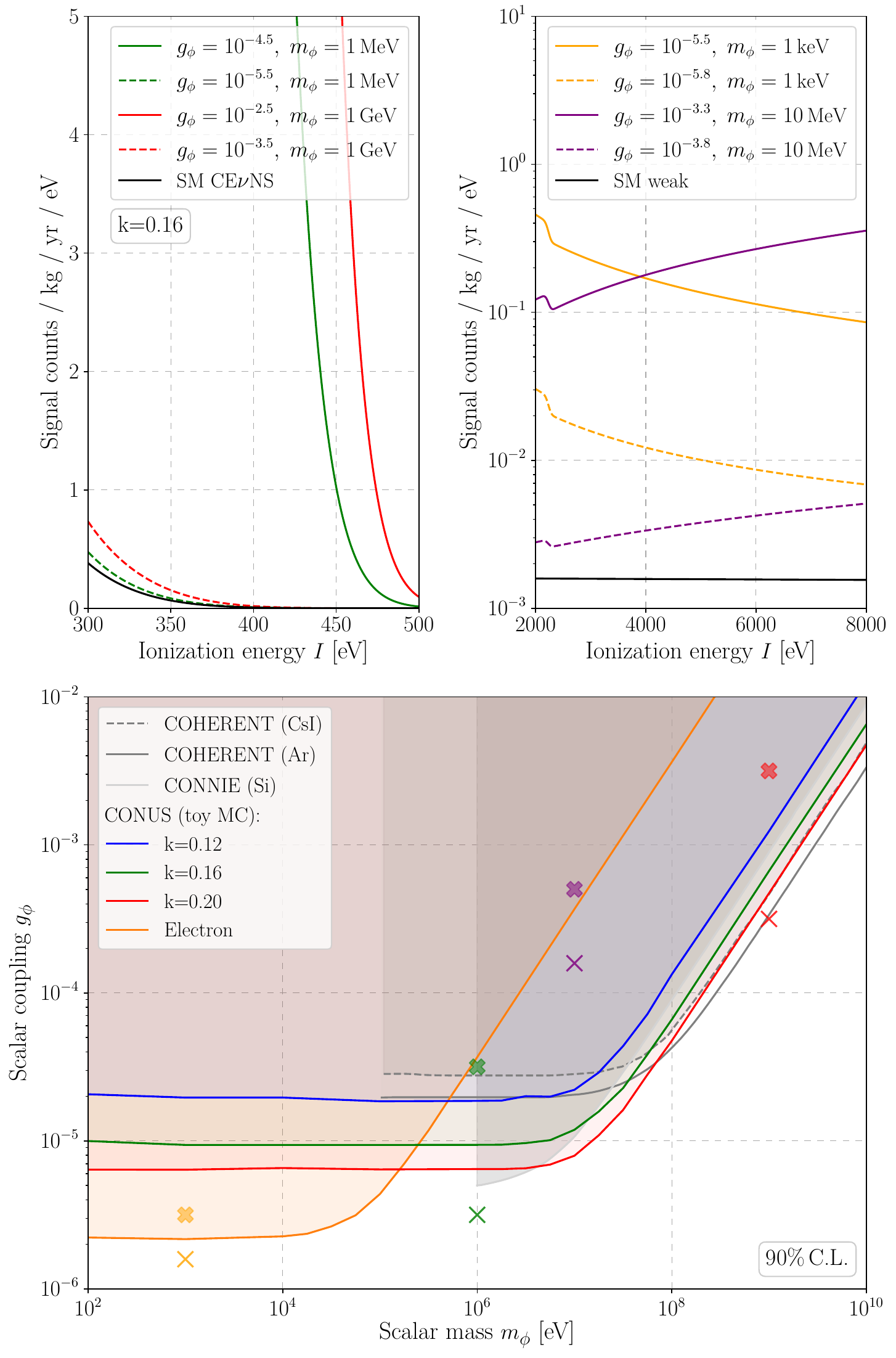}
	\caption{Top: Expected light scalar signals of detector C1 in the low energy region below $500 \eV_{ee}$ for a quenching parameter of $k = 0.16$ (left) and in the higher energy region between $2$ and $8 \keV_{ee}$ (right) for different couplings and masses in comparison to the SM signals of CE$\nu$NS and elastic neutrino-electron scattering, respectively.
    Note that the wiggles at $\sim2\keV$ are not artifacts but result from the applied reactor model.
	Bottom: Limits~(90\%\,C.L.) on the light scalar mediator parameters $(m_{\phi}, g_{\phi})$ deduced from CE$\nu$NS and neutrino-electron scattering with the \textsc{Run}-1 \textsc{Conus} data sets. 
	As before, we point out the exemplary parameter points of the signal spectra above.
	Bold crosses indicate parameter points that can already be excluded while regular crosses refer to points that are still in agreement with the data.
	For comparison, limits obtained from \textsc{Coherent} (CsI and Ar) data~(90\%\,C.L.)~\cite{Miranda:2020tif} and \textsc{Connie}~(95\%\,C.L.)~\cite{Aguilar-Arevalo:2019zme} are shown.
	}
	\label{fig:light_scalar_constraints}
\end{figure}

%%%%%%%%%%%%%%%%%%%%%%%%%%%%%%%%%%%%%%%%%%%%%%%%%%%%%%%%%%%%%%%%%%%%%%%%%%%%%%%%

Finally, we investigate elastic neutrino-nucleus and neutrino-electron scattering induced by a light scalar mediator $\phi$.
We select a simple benchmark model, i.e.\ a CP-even massive real scalar boson with pure scalar-type couplings to the first generation of leptons and quark. 
The Lagrangian of this simplified model is given by~\cite{Cerdeno:2016sfi}
\begin{align}\label{eq:lagrangian_light_scalar}
    \mathcal{L}_{\phi} = \phi\left( g^{q \mathrm{S}}_{\phi} \bar{q}q + g^{e \mathrm{S}}_{\phi} \bar{e}e + g^{\nu \mathrm{S}}_{\phi} \bar{\nu}_{R} \nu_{L}  + \text{h.c.} \right) 
    - \frac{1}{2} m^{2}_{\phi} \phi^2\, ,
\end{align}
with the individual scalar coupling $g^{x \mathrm{S}}_{\phi}$ ($x = \{\nu,\, e,\, q\}$ and $q = \{u,\, d\}$).
As for the vector mediator case, we put model-building aspects aside and work with this simplified model even though a realistic low-energy model needs to be more complex to become consistent with the SM symmetries \cite{Bell:2016ekl}.
Along the line of Refs.~\cite{Kosmas:2017tsq,Billard:2018jnl}, we also ignore resulting consequences for neutrino phenomenology in this analysis.

The associated neutrino-nucleus scattering cross section takes the form~\cite{Kosmas:2017tsq,Cerdeno:2016sfi}
\begin{align}\label{eq:cross_section_light_scalar_cenns}
    \left(\frac{d\sigma}{dT_{A}}\right)_{\rm CE\nu NS + \phi} = \left(\frac{d\sigma}{dT_{A}}\right)_{\rm CE\nu NS} 
    + \frac{(g^{\nu \mathrm{S}}_{\phi} \mathcal{Q}_{\phi})^{2} m_{A}^{2} T_{A} }{4\pi E_{\nu}^{2} (2 m_{A} T_{A}  + m_{\phi}^{2})^{2}}\, ,
\end{align}
with the nuclear charge associated to the light scalar mediator being~\cite{DelNobile:2013sia}
\begin{align}\label{eq:nuclear_charge_light_scalar}
    \mathcal{Q}_{\phi} = \sum_{N,q} g^{q \mathrm{S}}_{\phi} \frac{m_{N}}{m_{q}}f^{(N)}_{T,q} \rightarrow g_{\phi} (14N+15.1Z)\, .
\end{align}
The last step is obtained by assuming a universal coupling to leptons and quarks, and summing up all nucleon form factors $f_{T,q}^{(N)}$, which incorporate the effective low-energy couplings of the scalar $\phi$ to the nucleons $N=\{p,n\}$, cf.~Ref.~\cite{DelNobile:2013sia}.
Thus, with the assumption of a universal coupling, the corresponding part of the cross section in Eq.~\eqref{eq:cross_section_light_scalar_cenns} scales with $g_{\phi}^{4}$ and the model's parameter space is now spanned by only two parameters, the scalar mass $m_{\phi}$ and its couplings to fermions $g_{\phi}$.
Since the scalar-neutrino interaction flips chirality (in contrast to the chirality-conserving SM case), there is no interference and the scalar cross section is simply added to the SM CE$\nu$NS signal.
Another interesting aspect that appears in Eq.~\eqref{eq:cross_section_light_scalar_cenns} is the scaling with the recoil energy $T_{A}$ in comparison to the vector case, cf.~Eq.~\eqref{eq:cross_section_light_vector_cenns}.
For the scalar mediator, the corresponding part of the cross section scales with $1/T_{A}$, whereas in the vector case it scales with $1/T_{A}^2$, leading to a less steep signal.

The Lagrangian in Eq.~\eqref{eq:lagrangian_light_scalar} also induces an additional interaction between neutrinos and electrons.
Thus, there is an contribution to the cross section for neutrino-electron scattering, leading in total to~\cite{Cerdeno:2016sfi}
\begin{align}\label{eq:cross_section_light_scalar_nu_e}
    \left(\frac{d\sigma}{dT_{e}}\right)_{\nu e + \phi} = 
    \left(\frac{d\sigma}{dT}\right)_{\nu e} + \frac{ (g_{\phi}^{\nu \mathrm{S}} g^{e \mathrm{S}}_{\phi})^{2}\,  m_{e}^{2} T_{e} }{4\pi E_{\nu}^{2} (2 m_{e} T_{e}  + m_{\phi}^{2})^{2}}\, .
\end{align}
Under the assumption of universal scalar couplings, this shrinks down to the same quartic dependence as for neutrino-nucleus scattering, i.e. $(g_{\phi}^{\nu \mathrm{S}} g^{e \mathrm{S}}_{\phi})^{2} \rightarrow g_{\phi}^{4}$.
As for the case of a light vector mediator, the denominator in Eq.~\eqref{eq:cross_section_light_scalar_nu_e} can be separated into two different cases, i.e.\ $2m_{e}T_{e}\ll m_{\phi}^{2}$ and $2m_{e}T_{e}\gg m_{\phi}^{2}$, which correspond to the different behaviors of the obtained limit curves.

The expected event rates and the signal shape of elastic neutrino-nucleus and neutrino-electron scattering mediated by a light scalar are depicted in the upper left and right plots of Figure~\ref{fig:light_scalar_constraints}, respectively.
For comparison to the different signal expectations (two coupling values for each of the two scalar mediator masses), we also indicated the SM signal channels. 
By comparing the upper left plots of Figure~\ref{fig:light_vector_constraints} and Figure~\ref{fig:light_scalar_constraints}, one notes the previously mentioned difference in steepness or scaling with $T_{A}$ between the scalar and the vector mediator.
Further, this different scaling yields a different behavior for electron scatterings at higher energies, cf.~upper right plots of Figure~\ref{fig:light_vector_constraints} and Figure~\ref{fig:light_scalar_constraints}.
Here, the electron scattering exhibits a linear dependence on the recoil energy. 
In the end, this difference leads to stronger limits for the scalar mediator, which are displayed in the lower plot of Figure~\ref{fig:light_scalar_constraints}.
For comparison, we also show the limits obtained from \textsc{Coherent} and \textsc{Connie} and marked the parameter points of the upper plots with crosses.
Again, we highlighted both cases, points that are already excluded as well as points that still agree with the used data set. 
The lowest coupling value that can be probed with CE$\nu$NS is $\sim 10^{-5}$ for the currently most favoured quenching value of $k=0.16$, while elastic neutrino-electron scattering allows us to constrain the coupling down to $\sim 2\cdot10^{-6}$ for lowest mediator masses. 
As before, competitive CE$\nu$NS bounds can be gained for especially low mediator masses, i.e.\ below $\sim1\MeV$, which is attributed to the low neutrino energy provided by the reactor antineutrinos.

%%%%%%%%%%%%%%%%%%%%%%%%%%%%%%%%%%%%%%%%%%%%%%%%%%%%%%%%%%%%%%%%%%%%%%%%%%%%%%%%
%%%%%%%%%%%%%%%%%%%%%%%%%%%%%%%%%%%%%%%%%%%%%%%%%%%%%%%%%%%%%%%%%%%%%%%%%%%%%%%%

\section{Conclusions}\label{sec:conclusion}

\noindent The \textsc{Conus} experiment aims at the detection of CE$\nu$NS with four HPGe detectors in a sophisticated shield at 17.1\,m-distance to the 3.9\,GW$_{\rm th}$ core of the nuclear power plant in Brokdorf, Germany.
After a first spectral analysis devoted to the CE$\nu$NS search in \textsc{Conus} data, cf.~Ref.~\cite{CONUS:2020skt}, we used here \textsc{Run-1} data to constrain several BSM models. 
In particular, we searched for modifications of CE$\nu$NS due to NSIs of both tensor and vector type as well as light vector and scalar mediators.
The latter two have been tested as so-called simplified models on their impact on CE$\nu$NS and neutrino-electron scattering.
We make use of a similar analysis procedure that has already been used in the first CE$\nu$NS investigation, including all systematic uncertainties therein.
Small modifications have been applied due to the inclusion of uncertainties in the background MC simulation used in the higher energy spectrum, cf.~Section~\ref{sec:methods} and the background-related publication~\cite{CONUSbackground}.
Further, a refined noise edge parameterization was applied, leading to energy thresholds of the ROIs that are slightly lower compared to the analysis in Ref.~\cite{CONUS:2020skt}.

During our analysis, the likelihood function, cf.~Eq.~\eqref{eq:likelihood_full}, was varied with the cross sections of the individual models.
Limits were derived from data of three detectors in the experiment's first data collection period \textsc{Run-1}.
For the investigation of neutrino-electron scatterings above $2\keV_{ee}$, a data set with extended exposure is used to increase the experimental sensitivity, cf.~Table~\ref{tab:datasets} for an overview of all data sets used throughout this work.
For \textsc{Conus}, quenching, i.e.\ the fraction of nuclear recoil energy available as ionization for signal formation, is the least known input parameter and thus the dominating uncertainty.
In combination with neutrino energies below 10~MeV, this renders CE$\nu$NS measurements at a reactor site especially demanding.
Thus, we derive our BSM constraints for three different quenching parameters which span the range of currently favored values: $k=\{0.12, 0.16, 0.20\}$, where $k$ represents the quenching factor at recoil energies around $1 \keV_{ee}$, cf.~Section~\ref{sec:methods}.
The obtained bounds, except in the case of vector NSIs, are at least in some regions of the parameter space competitive with existing bounds from other CE$\nu$NS experiments, cf.~Section~\ref{sec:bsm_study}.
For tensor NSIs, we present limits that represent the world's best limits on electron-type couplings to up- and down-type quarks from CE$\nu$NS.
The scale of associated BSM physics can be constrained to lie above $\sim 360\GeV$, cf.~Figure~\ref{fig:tensor_nsi_bounds}.
Corresponding bounds in the case of vector-type NSIs are highly dependent on the quenching parameter $k$ and at the moment not competitive to existing bounds due to the limited sensitivity of \textsc{Conus} on the CE$\nu$NS signal itself, cf.~Figure~\ref{fig:vector_nsi_bounds}.
Since reactor antineutrinos are emitted at lower energies than neutrinos from a $\pi$-DAR source, our bounds on light scalar or vector mediators are stronger at smaller mediator masses. 
For higher masses, neutrinos from a $\pi$-DAR source yield currently the strongest CE$\nu$NS limits, cf.~Figure~\ref{fig:light_vector_constraints} and Figure~\ref{fig:light_scalar_constraints}.
Moreover, limits obtained from electron scatterings are stronger than the ones obtained from CE$\nu$NS for masses below $\sim 10\MeV$ and $\sim 1\MeV$ for vector and scalar mediators, respectively. 
However, we note that the shown parameter space region can only be excluded for models that incorporate electron and quark interactions with universal couplings.
For more specific frameworks, i.e.\ nucleophilic/leptophilic mediators or non-universal couplings, the obtained contours have to be viewed individually and/or with appropriate corrections. 

After a series of experimental improvements, i.e.\ an advanced data acquisition system and more stable environmental conditions, \textsc{Conus} continues data collection.
Thus, for the future we expect our bounds to strengthen with more exposure.
After the reactor shutdown at the end of 2021, additional OFF data are expected to increase the experimental sensitivity.
Further, the \textsc{Conus} Collaboration developed a program to pin down the dominating uncertainty related to the not well known quenching factor in germanium. 
Our recently conducted measurement is indicating a quenching factor value that agrees with the currently favored one and that follows the Lindhard theory down to nuclear recoils of a few keV, cf.~Ref.~\cite{CONUSquenching}.
With a future CE$\nu$NS detection via the \textsc{Conus} set-up, we expect stronger bounds, especially in the case of vector NSIs.
Then, investigation of further BSM topics like neutrino electromagnetic properties, sterile neutrino and dark matter will lead to further constraints.
An investigation of neutrino magnetic moments via neutrino-electron scattering at energies above 2\,keV$_{ee}$ can be found in Ref.~\cite{CONUS:2022qbb}.

While first BSM constraints of \textsc{Coherent}, \textsc{Connie}, \textsc{Conus} and \textsc{Ncc-1701} at \textsc{Dresden-II}~\cite{Colaresi:2021kus} already revealed the huge potential of CE$\nu$NS measurements, which can be viewed as a proof of principle by itself, more experiments are going to contribute further knowledge by using different target elements and detection technologies.
There are various endeavors close to nuclear reactors and first sensitivity studies for the European Spallation Source (\textsc{Ess}) already exist, cf.~Refs.~\cite{Blennow:2019bvl,Baxter:2019mcx}.
Taking advantage of these different neutrino sources, in terms of complementary measurements between reactor and neutrino beam experiments, allows for further interesting physics investigations~\cite{Dent:2017mpr,Lindner:2017nmb}.
Therefore, the next generation of CE$\nu$NS experiments promises an active field with new approaches and interesting possibilities~\cite{Bellenghi:2019vtc,Cadeddu:2019qmv,Abdullah:2020iiv,AristizabalSierra:2021uob,Fernandez-Moroni:2020yyl,Tomalak:2020zfh} and represents another step towards the era of precision neutrino physics.

%%%%%%%%%%%%%%%%%%%%%%%%%%%%%%%%%%%%%%%%%%%%%%%%%%%%%%%%%%%%%%%%%%%%%%%%%%%%%%%%
%%%%%%%%%%%%%%%%%%%%%%%%%%%%%%%%%%%%%%%%%%%%%%%%%%%%%%%%%%%%%%%%%%%%%%%%%%%%%%%%

\acknowledgments

We thank the technical and administrative staff who helped building the experiment, in particular the MPIK workshops and Mirion Technologies (Canberra) in Lingolsheim. 
We express our gratitude to the PreussenElektra GmbH for enduring and prolific support and for hosting the \textsc{Conus} experiment. 
Moreover, we thank Dr.~S.~Schoppmann (MPIK) for assistance on the analysis and Dr.~M.~Seidl (PreussenElektra GmbH) for providing simulation data on the fission rate evolution over a reactor cycle. 
Further thanks are directed to K.~Zink and the MPIK IT department for their support concerning the institute's cluster infrastructure.
We are grateful to G.~Vogt (MPIK) for providing an artist's view of our experiment (Figure 1).
The \textsc{Conus} experiment is supported financially by the Max Planck Society (MPG) and T.~Rink by the IMPRS-PTFS as well as by the German Research Foundation (DFG) through the research training group GRK 1940.

%%%%%%%%%%%%%%%%%%%%%%%%%%%%%%%%%%%%%%%%%%%%%%%%%%%%%%%%%%%%%%%%%%%%%%%%%%%%%%%%

\bibliography{literature} 

\providecommand{\href}[2]{#2}\begingroup\raggedright\begin{thebibliography}{100}

\bibitem{Freedman:1973yd}
D.~Z. Freedman, \emph{Coherent effects of a weak neutral current},
  \href{https://doi.org/10.1103/PhysRevD.9.1389}{\emph{Physical Review D}
  {\bfseries 9} (1974) 1389}.

\bibitem{Tubbs:1975jx}
D.~L. Tubbs and D.~N. Schramm, \emph{Neutrino {{Opacities}} at {{High
  Temperatures}} and {{Densities}}},
  \href{https://doi.org/10.1086/153909}{\emph{The Astrophysical Journal}
  {\bfseries 201} (1975) 467}.

\bibitem{Drukier:1983gj}
A.~Drukier and L.~Stodolsky, \emph{Principles and {{Applications}} of a
  {{Neutral Current Detector}} for {{Neutrino Physics}} and {{Astronomy}}},
  \href{https://doi.org/10.1103/PhysRevD.30.2295}{\emph{Phys.Rev.D} {\bfseries
  30} (1984) 2295}.

\bibitem{COHERENT:2017ipa}
{\scshape COHERENT} collaboration, D.~Akimov et~al., \emph{{Observation of
  Coherent Elastic Neutrino-Nucleus Scattering}},
  \href{https://doi.org/10.1126/science.aao0990}{\emph{Science} {\bfseries 357}
  (2017) 1123} [\href{https://arxiv.org/abs/1708.01294}{{\ttfamily
  1708.01294}}].

\bibitem{COHERENT:2020iec}
{\scshape COHERENT} collaboration, D.~Akimov et~al., \emph{{First Measurement
  of Coherent Elastic Neutrino-Nucleus Scattering on Argon}},
  \href{https://doi.org/10.1103/PhysRevLett.126.012002}{\emph{Phys. Rev. Lett.}
  {\bfseries 126} (2021) 012002}
  [\href{https://arxiv.org/abs/2003.10630}{{\ttfamily 2003.10630}}].

\bibitem{CONUS:2020skt}
{\scshape CONUS} collaboration, H.~Bonet et~al., \emph{{Constraints on Elastic
  Neutrino Nucleus Scattering in the Fully Coherent Regime from the CONUS
  Experiment}},
  \href{https://doi.org/10.1103/PhysRevLett.126.041804}{\emph{Phys. Rev. Lett.}
  {\bfseries 126} (2021) 041804}
  [\href{https://arxiv.org/abs/2011.00210}{{\ttfamily 2011.00210}}].

\bibitem{Kosmas:2017tsq}
D.~K. Papoulias and T.~S. Kosmas, \emph{{{COHERENT}} constraints to
  conventional and exotic neutrino physics},
  \href{https://doi.org/10.1103/PhysRevD.97.033003}{\emph{Phys.Rev.D}
  {\bfseries 97} (2018) } [\href{https://arxiv.org/abs/1711.09773}{{\ttfamily
  1711.09773}}].

\bibitem{Billard:2018jnl}
J.~Billard, J.~Johnston and B.~J. Kavanagh, \emph{Prospects for exploring {{New
  Physics}} in {{Coherent Elastic Neutrino}}-{{Nucleus Scattering}}},
  \href{https://doi.org/10.1088/1475-7516/2018/11/016}{\emph{JCAP} {\bfseries
  11} (2018) 016} [\href{https://arxiv.org/abs/1805.01798}{{\ttfamily
  1805.01798}}].

\bibitem{Khan:2019cvi}
A.~N. Khan and W.~Rodejohann, \emph{New physics from {{COHERENT}} data with an
  improved quenching factor},
  \href{https://doi.org/10.1103/PhysRevD.100.113003}{\emph{Phys.Rev.D}
  {\bfseries 100} (2019) } [\href{https://arxiv.org/abs/1907.12444}{{\ttfamily
  1907.12444}}].

\bibitem{Papoulias:2019txv}
D.~K. Papoulias, \emph{{{COHERENT}} constraints after the {{COHERENT}}-2020
  quenching factor measurement},
  \href{https://doi.org/10.1103/PhysRevD.102.113004}{\emph{Phys.Rev.D}
  {\bfseries 102} (2020) } [\href{https://arxiv.org/abs/1907.11644}{{\ttfamily
  1907.11644}}].

\bibitem{Freedman:1977xn}
D.~Z. Freedman, D.~N. Schramm and D.~L. Tubbs, \emph{The {{Weak Neutral
  Current}} and its {{Effects}} in {{Stellar Collapse}}},
  \href{https://doi.org/10.1146/annurev.ns.27.120177.001123}{\emph{Annual
  Review of Nuclear Science} {\bfseries 27} (1977) 167}.

\bibitem{Amanik:2006ad}
P.~S. Amanik and G.~M. Fuller, \emph{Stellar {{Collapse Dynamics With Neutrino
  Flavor Changing Neutral Currents}}},
  \href{https://doi.org/10.1103/PhysRevD.75.083008}{\emph{Phys.Rev.D}
  {\bfseries 75} (2007) }
  [\href{https://arxiv.org/abs/astro-ph/0606607}{{\ttfamily
  astro-ph/0606607}}].

\bibitem{Balasi:2015dba}
K.~G. Balasi, K.~Langanke and G.~{Mart{\'i}nez-Pinedo},
  \emph{Neutrino\textendash nucleus reactions and their role for supernova
  dynamics and nucleosynthesis},
  \href{https://doi.org/10.1016/j.ppnp.2015.08.001}{\emph{Prog.Part.Nucl.Phys.}
  {\bfseries 85} (2015) 33} [\href{https://arxiv.org/abs/1503.08095}{{\ttfamily
  1503.08095}}].

\bibitem{Amaya:2011sn}
M.~Biassoni and C.~Martinez, \emph{Study of supernova {$\nu$} -nucleus coherent
  scattering interactions},
  \href{https://doi.org/10.1016/j.astropartphys.2012.05.009}{\emph{Astropart.Phys.}
  {\bfseries 36} (2012) 151} [\href{https://arxiv.org/abs/1110.3536}{{\ttfamily
  1110.3536}}].

\bibitem{Brdar:2018zds}
V.~Brdar, M.~Lindner and X.-J. Xu, \emph{Neutrino astronomy with supernova
  neutrinos}, \href{https://doi.org/10.1088/1475-7516/2018/04/025}{\emph{JCAP}
  {\bfseries 04} (2018) 025}
  [\href{https://arxiv.org/abs/1802.02577}{{\ttfamily 1802.02577}}].

\bibitem{Agnes:2020pbw}
{\scshape DarkSide 20k} collaboration, P.~Agnes et~al., \emph{{Sensitivity of
  future liquid argon dark matter search experiments to core-collapse supernova
  neutrinos}}, \href{https://doi.org/10.1088/1475-7516/2021/03/043}{\emph{JCAP}
  {\bfseries 03} (2021) 043}
  [\href{https://arxiv.org/abs/2011.07819}{{\ttfamily 2011.07819}}].

\bibitem{Monroe:2007xp}
J.~Monroe and P.~Fisher, \emph{Neutrino {{Backgrounds}} to {{Dark Matter
  Searches}}},
  \href{https://doi.org/10.1103/PhysRevD.76.033007}{\emph{Phys.Rev.D}
  {\bfseries 76} (2007) } [\href{https://arxiv.org/abs/0706.3019}{{\ttfamily
  0706.3019}}].

\bibitem{Gutlein:2010tq}
A.~Gutlein, C.~Ciemniak, F.~{von Feilitzsch}, N.~Haag, M.~Hofmann, C.~Isaila
  et~al., \emph{Solar and atmospheric neutrinos: {{Background}} sources for the
  direct dark matter search},
  \href{https://doi.org/10.1016/j.astropartphys.2010.06.002}{\emph{Astropart.Phys.}
  {\bfseries 34} (2010) 90} [\href{https://arxiv.org/abs/1003.5530}{{\ttfamily
  1003.5530}}].

\bibitem{Cerdeno:2016sfi}
D.~G. Cerde\~no, M.~Fairbairn, T.~Jubb, P.~A.~N. Machado, A.~C. Vincent and
  C.~B\oe{}hm, \emph{{Physics from solar neutrinos in dark matter direct
  detection experiments}},
  \href{https://doi.org/10.1007/JHEP09(2016)048}{\emph{JHEP} {\bfseries 05}
  (2016) 118} [\href{https://arxiv.org/abs/1604.01025}{{\ttfamily
  1604.01025}}].

\bibitem{Bertuzzo:2017tuf}
E.~Bertuzzo, F.~F. Deppisch, S.~Kulkarni, Y.~F. Perez~Gonzalez and
  R.~Zukanovich~Funchal, \emph{Dark {{Matter}} and {{Exotic Neutrino
  Interactions}} in {{Direct Detection Searches}}},
  \href{https://doi.org/10.1007/JHEP04(2017)073}{\emph{JHEP} {\bfseries 04}
  (2017) 073} [\href{https://arxiv.org/abs/1701.07443}{{\ttfamily
  1701.07443}}].

\bibitem{Boehm:2018sux}
C.~B{\oe}hm, D.~G. Cerde{\~n}o, P.~a.~N. Machado, A.~{Olivares-Del Campo},
  E.~Perdomo and E.~Reid, \emph{How high is the neutrino floor?},
  \href{https://doi.org/10.1088/1475-7516/2019/01/043}{\emph{JCAP} {\bfseries
  01} (2019) 043} [\href{https://arxiv.org/abs/1809.06385}{{\ttfamily
  1809.06385}}].

\bibitem{Link:2019pbm}
J.~M. Link and X.-J. Xu, \emph{Searching for {{BSM}} neutrino interactions in
  dark matter detectors},
  \href{https://doi.org/10.1007/JHEP08(2019)004}{\emph{JHEP} {\bfseries 08}
  (2019) 004} [\href{https://arxiv.org/abs/1903.09891}{{\ttfamily
  1903.09891}}].

\bibitem{AristizabalSierra:2019ykk}
D.~Aristizabal~Sierra, B.~Dutta, S.~Liao and L.~E. Strigari, \emph{{Coherent
  elastic neutrino-nucleus scattering in multi-ton scale dark matter
  experiments: Classification of vector and scalar interactions new physics
  signals}}, \href{https://doi.org/10.1007/JHEP12(2019)124}{\emph{JHEP}
  {\bfseries 12} (2019) 124}
  [\href{https://arxiv.org/abs/1910.12437}{{\ttfamily 1910.12437}}].

\bibitem{Patton:2012jr}
K.~Patton, J.~Engel, G.~C. McLaughlin and N.~Schunck, \emph{Neutrino-nucleus
  coherent scattering as a probe of neutron density distributions},
  \href{https://doi.org/10.1103/PhysRevC.86.024612}{\emph{Phys.Rev.C}
  {\bfseries 86} (2012) } [\href{https://arxiv.org/abs/1207.0693}{{\ttfamily
  1207.0693}}].

\bibitem{Cadeddu:2017etk}
M.~Cadeddu, C.~Giunti, Y.~F. Li and Y.~Y. Zhang, \emph{{Average CsI neutron
  density distribution from COHERENT data}},
  \href{https://doi.org/10.1103/PhysRevLett.120.072501}{\emph{Phys. Rev. Lett.}
  {\bfseries 120} (2018) 072501}
  [\href{https://arxiv.org/abs/1710.02730}{{\ttfamily 1710.02730}}].

\bibitem{Coloma:2020nhf}
P.~Coloma, I.~Esteban, M.~C. {Gonzalez-Garcia} and J.~Menendez,
  \emph{Determining the nuclear neutron distribution from {{Coherent Elastic}}
  neutrino-{{Nucleus Scattering}}: Current results and future prospects},
  \href{https://doi.org/10.1007/JHEP08(2020)030}{\emph{JHEP} {\bfseries 08}
  (2020) 030} [\href{https://arxiv.org/abs/2006.08624}{{\ttfamily
  2006.08624}}].

\bibitem{VanDessel:2020epd}
N.~Van~Dessel, V.~Pandey, H.~Ray and N.~Jachowicz, \emph{{Nuclear Structure
  Physics in Coherent Elastic Neutrino-Nucleus Scattering}},
  \href{https://arxiv.org/abs/2007.03658}{{\ttfamily 2007.03658}}.

\bibitem{Lee:2015yht}
H.-S. Lee, \emph{{sin$^2 \theta_W$ theory and new physics}},
  \href{https://doi.org/10.3969/j.issn.0253-2778.2016.06.004}{\emph{J. Univ.
  Sci. Tech. China} {\bfseries 46} (2016) 470}
  [\href{https://arxiv.org/abs/1511.03783}{{\ttfamily 1511.03783}}].

\bibitem{Canas:2018rng}
B.~C. Ca{\~n}as, E.~A. Garc{\'e}s, O.~G. Miranda and A.~Parada, \emph{Future
  perspectives for a weak mixing angle measurement in coherent elastic neutrino
  nucleus scattering experiments},
  \href{https://doi.org/10.1016/j.physletb.2018.07.049}{\emph{Phys.Lett.B}
  {\bfseries 784} (2018) 159}
  [\href{https://arxiv.org/abs/1806.01310}{{\ttfamily 1806.01310}}].

\bibitem{Huang:2019ene}
X.-R. Huang and L.-W. Chen, \emph{{Neutron Skin in CsI and Low-Energy Effective
  Weak Mixing Angle from COHERENT Data}},
  \href{https://doi.org/10.1103/PhysRevD.100.071301}{\emph{Phys. Rev. D}
  {\bfseries 100} (2019) 071301}
  [\href{https://arxiv.org/abs/1902.07625}{{\ttfamily 1902.07625}}].

\bibitem{Fernandez-Moroni:2020yyl}
G.~{Fernandez-Moroni}, P.~A.~N. Machado, I.~{Martinez-Soler}, Y.~F.
  {Perez-Gonzalez}, D.~Rodrigues and S.~{Rosauro-Alcaraz}, \emph{The physics
  potential of a reactor neutrino experiment with {{Skipper CCDs}}:
  {{Measuring}} the weak mixing angle},
  \href{https://doi.org/10.1007/JHEP03(2021)186}{\emph{JHEP} {\bfseries 03}
  (2021) 186} [\href{https://arxiv.org/abs/2009.10741}{{\ttfamily
  2009.10741}}].

\bibitem{Barranco:2005yy}
J.~Barranco, O.~G. Miranda and T.~I. Rashba, \emph{Probing new physics with
  coherent neutrino scattering off nuclei},
  \href{https://doi.org/10.1088/1126-6708/2005/12/021}{\emph{JHEP} {\bfseries
  12} (2005) 021} [\href{https://arxiv.org/abs/hep-ph/0508299}{{\ttfamily
  hep-ph/0508299}}].

\bibitem{Barranco:2007tz}
J.~Barranco, O.~G. Miranda and T.~I. Rashba, \emph{Low energy neutrino
  experiments sensitivity to physics beyond the {{Standard Model}}},
  \href{https://doi.org/10.1103/PhysRevD.76.073008}{\emph{Phys.Rev.D}
  {\bfseries 76} (2007) }
  [\href{https://arxiv.org/abs/hep-ph/0702175}{{\ttfamily hep-ph/0702175}}].

\bibitem{Barranco:2011wx}
J.~Barranco, A.~Bolanos, E.~A. Garces, O.~G. Miranda and T.~I. Rashba,
  \emph{Tensorial {{NSI}} and {{Unparticle}} physics in neutrino scattering},
  \href{https://doi.org/10.1142/S0217751X12501473}{\emph{Int.J.Mod.Phys.A}
  {\bfseries 27} (2012) } [\href{https://arxiv.org/abs/1108.1220}{{\ttfamily
  1108.1220}}].

\bibitem{Lindner:2016wff}
M.~Lindner, W.~Rodejohann and X.-J. Xu, \emph{Coherent {{Neutrino}}-{{Nucleus
  Scattering}} and new {{Neutrino Interactions}}},
  \href{https://doi.org/10.1007/JHEP03(2017)097}{\emph{JHEP} {\bfseries 03}
  (2017) 097} [\href{https://arxiv.org/abs/1612.04150}{{\ttfamily
  1612.04150}}].

\bibitem{Coloma:2017ncl}
P.~Coloma, M.~C. Gonzalez-Garcia, M.~Maltoni and T.~Schwetz, \emph{{COHERENT
  Enlightenment of the Neutrino Dark Side}},
  \href{https://doi.org/10.1103/PhysRevD.96.115007}{\emph{Phys. Rev. D}
  {\bfseries 96} (2017) 115007}
  [\href{https://arxiv.org/abs/1708.02899}{{\ttfamily 1708.02899}}].

\bibitem{Liao:2017uzy}
J.~Liao and D.~Marfatia, \emph{{COHERENT constraints on nonstandard neutrino
  interactions}},
  \href{https://doi.org/10.1016/j.physletb.2017.10.046}{\emph{Phys. Lett. B}
  {\bfseries 775} (2017) 54}
  [\href{https://arxiv.org/abs/1708.04255}{{\ttfamily 1708.04255}}].

\bibitem{Bischer:2018zbd}
I.~Bischer, W.~Rodejohann and X.-J. Xu, \emph{Loop-induced {{Neutrino
  Non}}-{{Standard Interactions}}},
  \href{https://doi.org/10.1007/JHEP10(2018)096}{\emph{JHEP} {\bfseries 10}
  (2018) 096} [\href{https://arxiv.org/abs/1807.08102}{{\ttfamily
  1807.08102}}].

\bibitem{Dev:2019anc}
P.~S. Bhupal~Dev, K.~S. Babu, P.~B. Denton, P.~A.~N. Machado, C.~A.
  Arg{\"u}elles, J.~L. Barrow et~al., \emph{Neutrino {{Non}}-{{Standard
  Interactions}}: {{A Status Report}}},
  \href{https://doi.org/10.21468/SciPostPhysProc.2.001}{\emph{SciPost
  Phys.Proc.} {\bfseries 2} (2019) 001}
  [\href{https://arxiv.org/abs/1907.00991}{{\ttfamily 1907.00991}}].

\bibitem{Giunti:2019xpr}
C.~Giunti, \emph{{General COHERENT constraints on neutrino nonstandard
  interactions}},
  \href{https://doi.org/10.1103/PhysRevD.101.035039}{\emph{Phys. Rev. D}
  {\bfseries 101} (2020) 035039}
  [\href{https://arxiv.org/abs/1909.00466}{{\ttfamily 1909.00466}}].

\bibitem{Denton:2020hop}
P.~B. Denton and J.~Gehrlein, \emph{{A Statistical Analysis of the COHERENT
  Data and Applications to New Physics}},
  \href{https://doi.org/10.1007/JHEP04(2021)266}{\emph{JHEP} {\bfseries 04}
  (2021) 266} [\href{https://arxiv.org/abs/2008.06062}{{\ttfamily
  2008.06062}}].

\bibitem{Vogel:1989iv}
P.~Vogel and J.~Engel, \emph{Neutrino {{Electromagnetic Form}}-{{Factors}}},
  \href{https://doi.org/10.1103/PhysRevD.39.3378}{\emph{Phys.Rev.D} {\bfseries
  39} (1989) 3378}.

\bibitem{Giunti:2014ixa}
C.~Giunti and A.~Studenikin, \emph{Neutrino electromagnetic interactions: A
  window to new physics},
  \href{https://doi.org/10.1103/RevModPhys.87.531}{\emph{Rev.Mod.Phys.}
  {\bfseries 87} (2015) 531} [\href{https://arxiv.org/abs/1403.6344}{{\ttfamily
  1403.6344}}].

\bibitem{Cadeddu:2018dux}
M.~Cadeddu, C.~Giunti, K.~A. Kouzakov, Y.~F. Li, A.~I. Studenikin and Y.~Y.
  Zhang, \emph{{Neutrino Charge Radii from COHERENT Elastic Neutrino-Nucleus
  Scattering}}, \href{https://doi.org/10.1103/PhysRevD.98.113010}{\emph{Phys.
  Rev. D} {\bfseries 98} (2018) 113010}
  [\href{https://arxiv.org/abs/1810.05606}{{\ttfamily 1810.05606}}].

\bibitem{Miranda:2019wdy}
O.~G. Miranda, D.~K. Papoulias, M.~T{\'o}rtola and J.~W.~F. Valle,
  \emph{Probing neutrino transition magnetic moments with coherent elastic
  neutrino-nucleus scattering},
  \href{https://doi.org/10.1007/JHEP07(2019)103}{\emph{JHEP} {\bfseries 07}
  (2019) 103} [\href{https://arxiv.org/abs/1905.03750}{{\ttfamily
  1905.03750}}].

\bibitem{Cadeddu:2020lky}
M.~Cadeddu, F.~Dordei, C.~Giunti, Y.~F. Li, E.~Picciau and Y.~Y. Zhang,
  \emph{{Physics results from the first COHERENT observation of coherent
  elastic neutrino-nucleus scattering in argon and their combination with
  cesium-iodide data}},
  \href{https://doi.org/10.1103/PhysRevD.102.015030}{\emph{Phys. Rev. D}
  {\bfseries 102} (2020) 015030}
  [\href{https://arxiv.org/abs/2005.01645}{{\ttfamily 2005.01645}}].

\bibitem{deNiverville:2015mwa}
P.~{deNiverville}, M.~Pospelov and A.~Ritz, \emph{Light new physics in coherent
  neutrino-nucleus scattering experiments},
  \href{https://doi.org/10.1103/PhysRevD.92.095005}{\emph{Phys.Rev.D}
  {\bfseries 92} (2015) } [\href{https://arxiv.org/abs/1505.07805}{{\ttfamily
  1505.07805}}].

\bibitem{Dent:2016wcr}
J.~B. Dent, B.~Dutta, S.~Liao, J.~L. Newstead, L.~E. Strigari and J.~W. Walker,
  \emph{Probing light mediators at ultralow threshold energies with coherent
  elastic neutrino-nucleus scattering},
  \href{https://doi.org/10.1103/PhysRevD.96.095007}{\emph{Phys.Rev.D}
  {\bfseries 96} (2017) } [\href{https://arxiv.org/abs/1612.06350}{{\ttfamily
  1612.06350}}].

\bibitem{Farzan:2018gtr}
Y.~Farzan, M.~Lindner, W.~Rodejohann and X.-J. Xu, \emph{Probing neutrino
  coupling to a light scalar with coherent neutrino scattering},
  \href{https://doi.org/10.1007/JHEP05(2018)066}{\emph{JHEP} {\bfseries 05}
  (2018) 066} [\href{https://arxiv.org/abs/1802.05171}{{\ttfamily
  1802.05171}}].

\bibitem{Dent:2019ueq}
J.~B. Dent, B.~Dutta, D.~Kim, S.~Liao, R.~Mahapatra, K.~Sinha et~al., \emph{New
  {{Directions}} for {{Axion Searches}} via {{Scattering}} at {{Reactor
  Neutrino Experiments}}},
  \href{https://doi.org/10.1103/PhysRevLett.124.211804}{\emph{Phys.Rev.Lett.}
  {\bfseries 124} (2020) } [\href{https://arxiv.org/abs/1912.05733}{{\ttfamily
  1912.05733}}].

\bibitem{AristizabalSierra:2020rom}
D.~Aristizabal~Sierra, V.~De~Romeri, L.~J. Flores and D.~K. Papoulias,
  \emph{Axionlike particles searches in reactor experiments},
  \href{https://doi.org/10.1007/JHEP03(2021)294}{\emph{JHEP} {\bfseries 03}
  (2021) 294} [\href{https://arxiv.org/abs/2010.15712}{{\ttfamily
  2010.15712}}].

\bibitem{Dutta:2019eml}
B.~Dutta, S.~Liao, S.~Sinha and L.~E. Strigari, \emph{Searching for {{Beyond}}
  the {{Standard Model Physics}} with {{COHERENT Energy}} and {{Timing Data}}},
  \href{https://doi.org/10.1103/PhysRevLett.123.061801}{\emph{Phys.Rev.Lett.}
  {\bfseries 123} (2019) } [\href{https://arxiv.org/abs/1903.10666}{{\ttfamily
  1903.10666}}].

\bibitem{Aguilar-Arevalo:2019zme}
{\scshape CONNIE} collaboration, A.~Aguilar-Arevalo et~al., \emph{{Search for
  light mediators in the low-energy data of the CONNIE reactor neutrino
  experiment}}, \href{https://doi.org/10.1007/JHEP04(2020)054}{\emph{JHEP}
  {\bfseries 04} (2020) 054}
  [\href{https://arxiv.org/abs/1910.04951}{{\ttfamily 1910.04951}}].

\bibitem{Cadeddu:2020nbr}
M.~Cadeddu, N.~Cargioli, F.~Dordei, C.~Giunti, Y.~F. Li, E.~Picciau et~al.,
  \emph{Constraints on light vector mediators through coherent elastic neutrino
  nucleus scattering data from {{COHERENT}}},
  \href{https://doi.org/10.1007/JHEP01(2021)116}{\emph{JHEP} {\bfseries 01}
  (2021) 116} [\href{https://arxiv.org/abs/2008.05022}{{\ttfamily
  2008.05022}}].

\bibitem{Miranda:2020zji}
O.~G. Miranda, D.~K. Papoulias, M.~T\'ortola and J.~W.~F. Valle, \emph{{Probing
  new neutral gauge bosons with $CE\nu NS$ and neutrino-electron scattering}},
  \href{https://doi.org/10.1103/PhysRevD.101.073005}{\emph{Phys. Rev. D}
  {\bfseries 101} (2020) 073005}
  [\href{https://arxiv.org/abs/2002.01482}{{\ttfamily 2002.01482}}].

\bibitem{Brdar:2018qqj}
V.~Brdar, W.~Rodejohann and X.-J. Xu, \emph{Producing a new {{Fermion}} in
  {{Coherent Elastic Neutrino}}-{{Nucleus Scattering}}: From {{Neutrino Mass}}
  to {{Dark Matter}}},
  \href{https://doi.org/10.1007/JHEP12(2018)024}{\emph{JHEP} {\bfseries 12}
  (2018) 024} [\href{https://arxiv.org/abs/1810.03626}{{\ttfamily
  1810.03626}}].

\bibitem{Chang:2020jwl}
W.-F. Chang and J.~Liao, \emph{{Constraints on light singlet fermion
  interactions from coherent elastic neutrino-nucleus scattering}},
  \href{https://doi.org/10.1103/PhysRevD.102.075004}{\emph{Phys. Rev. D}
  {\bfseries 102} (2020) 075004}
  [\href{https://arxiv.org/abs/2002.10275}{{\ttfamily 2002.10275}}].

\bibitem{Aguilar:2001ty}
{\scshape LSND} collaboration, A.~Aguilar-Arevalo et~al., \emph{{Evidence for
  neutrino oscillations from the observation of $\bar{\nu}_e$ appearance in a
  $\bar{\nu}_\mu$ beam}},
  \href{https://doi.org/10.1103/PhysRevD.64.112007}{\emph{Phys. Rev. D}
  {\bfseries 64} (2001) 112007}
  [\href{https://arxiv.org/abs/hep-ex/0104049}{{\ttfamily hep-ex/0104049}}].

\bibitem{Mention:2011rk}
G.~Mention, M.~Fechner, T.~Lasserre, T.~A. Mueller, D.~Lhuillier, M.~Cribier
  et~al., \emph{The {{Reactor Antineutrino Anomaly}}},
  \href{https://doi.org/10.1103/PhysRevD.83.073006}{\emph{Phys.Rev.D}
  {\bfseries 83} (2011) } [\href{https://arxiv.org/abs/1101.2755}{{\ttfamily
  1101.2755}}].

\bibitem{Boser:2019rta}
S.~B{\"o}ser, C.~Buck, C.~Giunti, J.~Lesgourgues, L.~Ludhova, S.~Mertens
  et~al., \emph{Status of {{Light Sterile Neutrino Searches}}},
  \href{https://doi.org/10.1016/j.ppnp.2019.103736}{\emph{Prog.Part.Nucl.Phys.}
  {\bfseries 111} (2020) } [\href{https://arxiv.org/abs/1906.01739}{{\ttfamily
  1906.01739}}].

\bibitem{Formaggio:2011jt}
J.~A. Formaggio, E.~{Figueroa-Feliciano} and A.~J. Anderson, \emph{Sterile
  {{Neutrinos}}, {{Coherent Scattering}} and {{Oscillometry Measurements}} with
  {{Low}}-temperature {{Bolometers}}},
  \href{https://doi.org/10.1103/PhysRevD.85.013009}{\emph{Phys.Rev.D}
  {\bfseries 85} (2012) } [\href{https://arxiv.org/abs/1107.3512}{{\ttfamily
  1107.3512}}].

\bibitem{Dutta:2015nlo}
B.~Dutta, Y.~Gao, R.~Mahapatra, N.~Mirabolfathi, L.~E. Strigari and J.~W.
  Walker, \emph{Sensitivity to oscillation with a sterile fourth generation
  neutrino from ultra-low threshold neutrino-nucleus coherent scattering},
  \href{https://doi.org/10.1103/PhysRevD.94.093002}{\emph{Phys.Rev.D}
  {\bfseries 94} (2016) } [\href{https://arxiv.org/abs/1511.02834}{{\ttfamily
  1511.02834}}].

\bibitem{Canas:2017umu}
B.~C. Ca{\~n}as, E.~A. Garc{\'e}s, O.~G. Miranda and A.~Parada, \emph{The
  reactor antineutrino anomaly and low energy threshold neutrino experiments},
  \href{https://doi.org/10.1016/j.physletb.2017.11.074}{\emph{Phys.Lett.B}
  {\bfseries 776} (2018) 451}
  [\href{https://arxiv.org/abs/1708.09518}{{\ttfamily 1708.09518}}].

\bibitem{Blanco:2019vyp}
C.~Blanco, D.~Hooper and P.~Machado, \emph{Constraining {{Sterile Neutrino
  Interpretations}} of the {{LSND}} and {{MiniBooNE Anomalies}} with {{Coherent
  Neutrino Scattering Experiments}}},
  \href{https://doi.org/10.1103/PhysRevD.101.075051}{\emph{Phys.Rev.D}
  {\bfseries 101} (2020) } [\href{https://arxiv.org/abs/1901.08094}{{\ttfamily
  1901.08094}}].

\bibitem{Miranda:2020syh}
O.~G. Miranda, D.~K. Papoulias, O.~Sanders, M.~T{\'o}rtola and J.~W.~F. Valle,
  \emph{Future {{CEvNS}} experiments as probes of lepton unitarity and
  light-sterile neutrinos},
  \href{https://doi.org/10.1103/PhysRevD.102.113014}{\emph{Phys.Rev.D}
  {\bfseries 102} (2020) } [\href{https://arxiv.org/abs/2008.02759}{{\ttfamily
  2008.02759}}].

\bibitem{Hagmann:2004uv}
C.~Hagmann and A.~Bernstein, \emph{Two-phase emission detector for measuring
  coherent neutrino-nucleus scattering},
  \href{https://doi.org/10.1109/TNS.2004.836061}{\emph{IEEE Trans.Nucl.Sci.}
  {\bfseries 51} (2004) 2151}
  [\href{https://arxiv.org/abs/nucl-ex/0411004}{{\ttfamily nucl-ex/0411004}}].

\bibitem{Bernstein:2019hix}
A.~Bernstein, N.~Bowden, B.~L. Goldblum, P.~Huber, I.~Jovanovic and
  J.~Mattingly, \emph{Colloquium: {{Neutrino}} detectors as tools for nuclear
  security},
  \href{https://doi.org/10.1103/RevModPhys.92.011003}{\emph{Rev.Mod.Phys.}
  {\bfseries 92} (2020) } [\href{https://arxiv.org/abs/1908.07113}{{\ttfamily
  1908.07113}}].

\bibitem{Bowen:2020unj}
M.~Bowen and P.~Huber, \emph{Reactor neutrino applications and coherent elastic
  neutrino nucleus scattering},
  \href{https://doi.org/10.1103/PhysRevD.102.053008}{\emph{Phys.Rev.D}
  {\bfseries 102} (2020) } [\href{https://arxiv.org/abs/2005.10907}{{\ttfamily
  2005.10907}}].

\bibitem{Aguilar-Arevalo:2016khx}
{\scshape CONNIE} collaboration, A.~Aguilar-Arevalo et~al., \emph{{The CONNIE
  experiment}}, \href{https://doi.org/10.1088/1742-6596/761/1/012057}{\emph{J.
  Phys. Conf. Ser.} {\bfseries 761} (2016) 012057}
  [\href{https://arxiv.org/abs/1608.01565}{{\ttfamily 1608.01565}}].

\bibitem{Agnolet:2016zir}
{\scshape MINER} collaboration, G.~Agnolet et~al., \emph{{Background Studies
  for the MINER Coherent Neutrino Scattering Reactor Experiment}},
  \href{https://doi.org/10.1016/j.nima.2017.02.024}{\emph{Nucl. Instrum. Meth.
  A} {\bfseries 853} (2017) 53}
  [\href{https://arxiv.org/abs/1609.02066}{{\ttfamily 1609.02066}}].

\bibitem{Colaresi:2021kus}
J.~Colaresi, J.~I. Collar, T.~W. Hossbach, A.~R.~L. Kavner, C.~M. Lewis, A.~E.
  Robinson et~al., \emph{{First results from a search for coherent elastic
  neutrino-nucleus scattering at a reactor site}},
  \href{https://doi.org/10.1103/PhysRevD.104.072003}{\emph{Phys. Rev. D}
  {\bfseries 104} (2021) 072003}
  [\href{https://arxiv.org/abs/2108.02880}{{\ttfamily 2108.02880}}].

\bibitem{Choi:2020gkm}
J.~J. Choi, \emph{{Neutrino Elastic-scattering Observation with
  NaI[Tl](NEON)}}, \href{https://doi.org/10.22323/1.369.0047}{\emph{PoS}
  {\bfseries NuFact2019} (2020) 047}.

\bibitem{Strauss:2017cuu}
{\scshape {$\nu$}-cleus} collaboration, R.~Strauss et~al., \emph{{The
  $\nu$-cleus experiment: A gram-scale fiducial-volume cryogenic detector for
  the first detection of coherent neutrino-nucleus scattering}},
  \href{https://doi.org/10.1140/epjc/s10052-017-5068-2}{\emph{Eur. Phys. J. C}
  {\bfseries 77} (2017) 506}
  [\href{https://arxiv.org/abs/1704.04320}{{\ttfamily 1704.04320}}].

\bibitem{Belov:2015ufh}
{\scshape \ensuremath{\nu}GeN} collaboration, V.~Belov et~al., \emph{{The
  \ensuremath{\nu}GeN experiment at the Kalinin Nuclear Power Plant}},
  \href{https://doi.org/10.1088/1748-0221/10/12/P12011}{\emph{JINST} {\bfseries
  10} (2015) P12011}.

\bibitem{Akimov:2019ogx}
{\scshape RED-100} collaboration, D.~Y. Akimov et~al., \emph{{First
  ground-level laboratory test of the two-phase xenon emission detector
  RED-100}}, \href{https://doi.org/10.1088/1748-0221/15/02/P02020}{\emph{JINST}
  {\bfseries 15} (2020) P02020}
  [\href{https://arxiv.org/abs/1910.06190}{{\ttfamily 1910.06190}}].

\bibitem{Billard:2016giu}
{\scshape Ricochet} collaboration, J.~Billard et~al., \emph{{Coherent Neutrino
  Scattering with Low Temperature Bolometers at Chooz Reactor Complex}},
  \href{https://doi.org/10.1088/1361-6471/aa83d0}{\emph{J. Phys. G} {\bfseries
  44} (2017) 105101} [\href{https://arxiv.org/abs/1612.09035}{{\ttfamily
  1612.09035}}].

\bibitem{Wong:2004ru}
{\scshape TEXONO} collaboration, H.~T. Wong, \emph{{The TEXONO research program
  on neutrino and astroparticle physics}},
  \href{https://doi.org/10.1142/S0217732304014574}{\emph{Mod. Phys. Lett. A}
  {\bfseries 19} (2004) 1207}.

\bibitem{FernandezMoroni:2014qlq}
G.~Fernandez~Moroni, J.~Estrada, E.~E. Paolini, G.~Cancelo, J.~Tiffenberg and
  J.~Molina, \emph{{Charge Coupled Devices for detection of coherent
  neutrino-nucleus scattering}},
  \href{https://doi.org/10.1103/PhysRevD.91.072001}{\emph{Phys. Rev. D}
  {\bfseries 91} (2015) 072001}
  [\href{https://arxiv.org/abs/1405.5761}{{\ttfamily 1405.5761}}].

\bibitem{Strauss:2017cam}
{\scshape {$\nu$}-cleus} collaboration, R.~Strauss et~al., \emph{{Gram-scale
  cryogenic calorimeters for rare-event searches}},
  \href{https://doi.org/10.1103/PhysRevD.96.022009}{\emph{Phys. Rev. D}
  {\bfseries 96} (2017) 022009}
  [\href{https://arxiv.org/abs/1704.04317}{{\ttfamily 1704.04317}}].

\bibitem{Bonet:2020ntx}
{\scshape CONUS} collaboration, H.~Bonet et~al., \emph{{Large-size sub-keV
  sensitive germanium detectors for the CONUS experiment}},
  \href{https://doi.org/10.1140/epjc/s10052-021-09038-3}{\emph{Eur. Phys. J. C}
  {\bfseries 81} (2021) 267}
  [\href{https://arxiv.org/abs/2010.11241}{{\ttfamily 2010.11241}}].

\bibitem{Chepel:2012sj}
V.~Chepel and H.~Araujo, \emph{Liquid noble gas detectors for low energy
  particle physics},
  \href{https://doi.org/10.1088/1748-0221/8/04/R04001}{\emph{JINST} {\bfseries
  8} (2013) } [\href{https://arxiv.org/abs/1207.2292}{{\ttfamily 1207.2292}}].

\bibitem{Choi:2020qcj}
J.~J. Choi, B.~J. Park, C.~Ha, K.~W. Kim, S.~K. Kim, Y.~D. Kim et~al.,
  \emph{{Improving the light collection using a new NaI(Tl) crystal
  encapsulation}},
  \href{https://doi.org/10.1016/j.nima.2020.164556}{\emph{Nucl. Instrum. Meth.
  A} {\bfseries 981} (2020) 164556}
  [\href{https://arxiv.org/abs/2006.02573}{{\ttfamily 2006.02573}}].

\bibitem{Hayes:2016qnu}
A.~C. Hayes and P.~Vogel, \emph{Reactor {{Neutrino Spectra}}},
  \href{https://doi.org/10.1146/annurev-nucl-102115-044826}{\emph{Ann.Rev.Nucl.Part.Sci.}
  {\bfseries 66} (2016) 219}
  [\href{https://arxiv.org/abs/1605.02047}{{\ttfamily 1605.02047}}].

\bibitem{Huber:2011wv}
P.~Huber, \emph{{On the determination of anti-neutrino spectra from nuclear
  reactors}}, \href{https://doi.org/10.1103/PhysRevC.85.029901}{\emph{Phys.
  Rev. C} {\bfseries 84} (2011) 024617}
  [\href{https://arxiv.org/abs/1106.0687}{{\ttfamily 1106.0687}}].

\bibitem{Mueller:2011nm}
T.~A. Mueller, D.~Lhuillier, M.~Fallot, A.~Letourneau, S.~Cormon, M.~Fechner
  et~al., \emph{Improved {{Predictions}} of {{Reactor Antineutrino Spectra}}},
  \href{https://doi.org/10.1103/PhysRevC.83.054615}{\emph{Phys.Rev.C}
  {\bfseries 83} (2011) } [\href{https://arxiv.org/abs/1101.2663}{{\ttfamily
  1101.2663}}].

\bibitem{An:2016srz}
{\scshape Daya Bay} collaboration, F.~P. An et~al., \emph{{Improved Measurement
  of the Reactor Antineutrino Flux and Spectrum at Daya Bay}},
  \href{https://doi.org/10.1088/1674-1137/41/1/013002}{\emph{Chin. Phys. C}
  {\bfseries 41} (2017) 013002}
  [\href{https://arxiv.org/abs/1607.05378}{{\ttfamily 1607.05378}}].

\bibitem{Kopeikin:2003gu}
V.~Kopeikin, L.~Mikaelyan and V.~Sinev, \emph{Components of anti-neutrino
  emission in nuclear reactor},
  \href{https://doi.org/10.1134/1.1825513}{\emph{Phys.Atom.Nucl.} {\bfseries
  67} (2004) 1963} [\href{https://arxiv.org/abs/hep-ph/0308186}{{\ttfamily
  hep-ph/0308186}}].

\bibitem{Beda:2007hf}
{\scshape GEMMA} collaboration, A.~G. Beda, V.~B. Brudanin, E.~V. Demidova,
  V.~G. Egorov, M.~G. Gavrilov, M.~V. Shirchenko et~al., \emph{{First Result
  for Neutrino Magnetic Moment from Measurements with the GEMMA Spectrometer}},
  \href{https://doi.org/10.1134/S1063778807110063}{\emph{Phys. Atom. Nucl.}
  {\bfseries 70} (2007) 1873}
  [\href{https://arxiv.org/abs/0705.4576}{{\ttfamily 0705.4576}}].

\bibitem{Ma:2012bm}
X.~B. Ma, W.~L. Zhong, L.~Z. Wang, Y.~X. Chen and J.~Cao, \emph{Improved
  calculation of the energy release in neutron-induced fission},
  \href{https://doi.org/10.1103/PhysRevC.88.014605}{\emph{Phys.Rev.C}
  {\bfseries 88} (2013) } [\href{https://arxiv.org/abs/1212.6625}{{\ttfamily
  1212.6625}}].

\bibitem{Heusser:1995wd}
G.~Heusser, \emph{Low-radioactivity background techniques},
  \href{https://doi.org/10.1146/annurev.ns.45.120195.002551}{\emph{Ann.Rev.Nucl.Part.Sci.}
  {\bfseries 45} (1995) 543}.

\bibitem{Heusser:2015ifa}
G.~Heusser, M.~Weber, J.~Hakenm{\"u}ller, M.~Laubenstein, M.~Lindner,
  W.~Maneschg et~al., \emph{{{GIOVE}} - {{A}} new detector setup for high
  sensitivity germanium spectroscopy at shallow depth},
  \href{https://doi.org/10.1140/epjc/s10052-015-3704-2}{\emph{Eur.Phys.J.C}
  {\bfseries 75} (2015) 531}
  [\href{https://arxiv.org/abs/1507.03319}{{\ttfamily 1507.03319}}].

\bibitem{Hakenmuller:2019ecb}
{\scshape CONUS} collaboration, J.~Hakenm\"uller et~al., \emph{{Neutron-induced
  background in the CONUS experiment}},
  \href{https://doi.org/10.1140/epjc/s10052-019-7160-2}{\emph{Eur. Phys. J. C}
  {\bfseries 79} (2019) 699}
  [\href{https://arxiv.org/abs/1903.09269}{{\ttfamily 1903.09269}}].

\bibitem{CONUSbackground}
H.~Bonet et~al., \emph{{Full background decomposition of the CONUS
  experiment}},  \href{https://arxiv.org/abs/2112.09585}{{\ttfamily
  2112.09585}}.

\bibitem{Lindhard:1961zz}
J.~Lindhard and M.~Scharff, \emph{Energy {{Dissipation}} by {{Ions}} in the kev
  {{Region}}}, \href{https://doi.org/10.1103/PhysRev.124.128}{\emph{Phys.Rev.}
  {\bfseries 124} (1961) 128}.

\bibitem{Scholz:2016qos}
B.~J. Scholz, A.~E. Chavarria, J.~I. Collar, P.~Privitera and A.~E. Robinson,
  \emph{{Measurement of the low-energy quenching factor in germanium using an
  $^{88}$Y/Be photoneutron source}},
  \href{https://doi.org/10.1103/PhysRevD.94.122003}{\emph{Phys. Rev. D}
  {\bfseries 94} (2016) 122003}
  [\href{https://arxiv.org/abs/1608.03588}{{\ttfamily 1608.03588}}].

\bibitem{Collar:2021fcl}
J.~I. Collar, A.~R.~L. Kavner and C.~M. Lewis, \emph{{Germanium response to
  sub-keV nuclear recoils: a multipronged experimental characterization}},
  \href{https://doi.org/10.1103/PhysRevD.103.122003}{\emph{Phys. Rev. D}
  {\bfseries 103} (2021) 122003}
  [\href{https://arxiv.org/abs/2102.10089}{{\ttfamily 2102.10089}}].

\bibitem{Sorensen:2014sla}
P.~Sorensen, \emph{{Atomic limits in the search for galactic dark matter}},
  \href{https://doi.org/10.1103/PhysRevD.91.083509}{\emph{Phys. Rev. D}
  {\bfseries 91} (2015) 083509}
  [\href{https://arxiv.org/abs/1412.3028}{{\ttfamily 1412.3028}}].

\bibitem{Liao:2021yog}
J.~Liao, H.~Liu and D.~Marfatia, \emph{{Coherent neutrino scattering and the
  Migdal effect on the quenching factor}},
  \href{https://doi.org/10.1103/PhysRevD.104.015005}{\emph{Phys. Rev. D}
  {\bfseries 104} (2021) 015005}
  [\href{https://arxiv.org/abs/2104.01811}{{\ttfamily 2104.01811}}].

\bibitem{Wilks:1938dza}
S.~S. Wilks, \emph{The {{Large}}-{{Sample Distribution}} of the {{Likelihood
  Ratio}} for {{Testing Composite Hypotheses}}},
  \href{https://doi.org/10.1214/aoms/1177732360}{\emph{Annals Math.Statist.}
  {\bfseries 9} (1938) 60}.

\bibitem{Cowan:2010js}
G.~Cowan, K.~Cranmer, E.~Gross and O.~Vitells, \emph{Asymptotic formulae for
  likelihood-based tests of new physics},
  \href{https://doi.org/10.1140/epjc/s10052-011-1554-0}{\emph{Eur.Phys.J.C}
  {\bfseries 71} (2011) 1554}
  [\href{https://arxiv.org/abs/1007.1727}{{\ttfamily 1007.1727}}].

\bibitem{Lista:2016chp}
L.~Lista, \emph{{Practical Statistics for Particle Physicists}},  in
  \emph{{2016 European School of High-Energy Physics}}, pp.~213--258, 2017,
  \href{https://arxiv.org/abs/1609.04150}{{\ttfamily 1609.04150}},
  \href{https://doi.org/10.23730/CYRSP-2017-005.213}{DOI}.

\bibitem{Helm:1956zz}
R.~H. Helm, \emph{{Inelastic and Elastic Scattering of 187-MeV Electrons from
  Selected Even-Even Nuclei}},
  \href{https://doi.org/10.1103/PhysRev.104.1466}{\emph{Phys. Rev.} {\bfseries
  104} (1956) 1466}.

\bibitem{Klein:1999qj}
S.~Klein and J.~Nystrand, \emph{Exclusive vector meson production in
  relativistic heavy ion collisions},
  \href{https://doi.org/10.1103/PhysRevC.60.014903}{\emph{Phys.Rev.C}
  {\bfseries 60} (1999) }
  [\href{https://arxiv.org/abs/hep-ph/9902259}{{\ttfamily hep-ph/9902259}}].

\bibitem{COHERENT:2018gft}
{\scshape COHERENT} collaboration, D.~Akimov et~al., \emph{{COHERENT 2018 at
  the Spallation Neutron Source}},
  \href{https://arxiv.org/abs/1803.09183}{{\ttfamily 1803.09183}}.

\bibitem{AristizabalSierra:2019zmy}
D.~Aristizabal~Sierra, J.~Liao and D.~Marfatia, \emph{{Impact of form factor
  uncertainties on interpretations of coherent elastic neutrino-nucleus
  scattering data}}, \href{https://doi.org/10.1007/JHEP06(2019)141}{\emph{JHEP}
  {\bfseries 06} (2019) 141}
  [\href{https://arxiv.org/abs/1902.07398}{{\ttfamily 1902.07398}}].

\bibitem{Giunti:2007ry}
C.~Giunti and C.~W. Kim, \emph{{Fundamentals of Neutrino Physics and
  Astrophysics}}. Oxford Univ., 2007.

\bibitem{Mikaelyan:2002nv}
L.~A. Mikaelyan, \emph{Investigation of neutrino properties in experiments at
  nuclear reactors: {{Present}} status and prospects},
  \href{https://doi.org/10.1134/1.1495017}{\emph{Phys.Atom.Nucl.} {\bfseries
  65} (2002) 1173} [\href{https://arxiv.org/abs/hep-ph/0210047}{{\ttfamily
  hep-ph/0210047}}].

\bibitem{Perkins:1991}
S.~Perkins, \emph{Tables and Graphs of Atomic Subshell and Relaxation Data
  Derived from the LLNL Evaluated Atomic Data Library (EADL), Z}. Lawrence
  Livermore National Laboratory, 1991.

\bibitem{Jones:1971ya}
K.~W. Jones and H.~W. Kraner, \emph{{Stopping of 1- to 1.8-keV Ge-73 Atoms in
  Germanium}}, \href{https://doi.org/10.1103/PhysRevC.4.125}{\emph{Phys. Rev.
  C} {\bfseries 4} (1971) 125}.

\bibitem{Jones:1975zze}
K.~W. Jones and H.~W. Kraner, \emph{{Energy lost to ionization by 254-eV Ge-73
  atoms stopping in Ge}},
  \href{https://doi.org/10.1103/PhysRevA.11.1347}{\emph{Phys. Rev. A}
  {\bfseries 11} (1975) 1347}.

\bibitem{Messous:1995dn}
Y.~Messous, \emph{{Calibration of a Ge crystal with nuclear recoils for the
  development of a dark matter detector}},
  \href{https://doi.org/10.1016/0927-6505(95)00007-4}{\emph{Astropart. Phys.}
  {\bfseries 3} (1995) 361}.

\bibitem{Barbeau:2007qi}
P.~S. Barbeau, J.~I. Collar and O.~Tench, \emph{{Large-Mass Ultra-Low Noise
  Germanium Detectors: Performance and Applications in Neutrino and
  Astroparticle Physics}},
  \href{https://doi.org/10.1088/1475-7516/2007/09/009}{\emph{JCAP} {\bfseries
  09} (2007) 009} [\href{https://arxiv.org/abs/nucl-ex/0701012}{{\ttfamily
  nucl-ex/0701012}}].

\bibitem{Barker:2012ek}
D.~Barker and D.~M. Mei, \emph{{Germanium Detector Response to Nuclear Recoils
  in Searching for Dark Matter}},
  \href{https://doi.org/10.1016/j.astropartphys.2012.08.006}{\emph{Astropart.
  Phys.} {\bfseries 38} (2012) 1}
  [\href{https://arxiv.org/abs/1203.4620}{{\ttfamily 1203.4620}}].

\bibitem{CONUS:2022qbb}
{\scshape CONUS} collaboration, H.~Bonet et~al., \emph{{First limits on
  neutrino electromagnetic properties from the CONUS experiment}},
  \href{https://arxiv.org/abs/2201.12257}{{\ttfamily 2201.12257}}.

\bibitem{iminuit2020}
H.~Dembinski, P.~Ongmongkolkul, C.~Deil, D.~M. Hurtado, H.~Schreiner,
  M.~Feickert et~al., \emph{scikit-hep/iminuit: v2.0.0},  Dec., 2020.
\newblock 10.5281/zenodo.4310361.

\bibitem{minuit1975}
F.~James and M.~Roos, \emph{Minuit - a system for function minimization and
  analysis of the parameter errors and correlations},
  \href{https://doi.org/10.1016/0010-4655(75)90039-9}{\emph{Computer Physics
  Communications} {\bfseries 10} (1975) 343}.

\bibitem{ScientificComputing2007}
T.~E. Oliphant, \emph{Python for scientific computing},
  \href{https://doi.org/10.1109/MCSE.2007.58}{\emph{Computing in Science
  Engineering} {\bfseries 9} (2007) 10}.

\bibitem{ScientificComputing2011}
K.~J. Millman and M.~Aivazis, \emph{Python for scientists and engineers},
  \href{https://doi.org/10.1109/MCSE.2011.36}{\emph{Computing in Science
  Engineering} {\bfseries 13} (2011) 9}.

\bibitem{Scipy2020}
P.~Virtanen et~al., \emph{{{SciPy}} 1.0: Fundamental algorithms for scientific
  computing in {{Python}}},
  \href{https://doi.org/10.1038/s41592-019-0686-2}{\emph{Nature Methods}
  {\bfseries 17} (2020) 261}.

\bibitem{Matplotlib2007}
J.~D. Hunter, \emph{Matplotlib: A 2d graphics environment},
  \href{https://doi.org/10.1109/MCSE.2007.55}{\emph{Computing in Science
  Engineering} {\bfseries 9} (2007) 90}.

\bibitem{Ipython2007}
F.~Perez and B.~E. Granger, \emph{Ipython: A system for interactive scientific
  computing}, \href{https://doi.org/10.1109/MCSE.2007.53}{\emph{Computing in
  Science Engineering} {\bfseries 9} (2007) 21}.

\bibitem{Numpy2020}
C.~R. Harris et~al., \emph{Array programming with {{NumPy}}},
  \href{https://doi.org/10.1038/s41586-020-2649-2}{\emph{Nature} {\bfseries
  585} (2020) 357}.

\bibitem{Pandas2010}
W.~McKinney, \emph{Data {{Structures}} for {{Statistical Computing}} in
  {{Python}}},
  \href{https://doi.org/10.25080/Majora-92bf1922-00a}{\emph{Proceedings of the
  9th Python in Science Conference} (2010) 56}.

\bibitem{JupyterLab}
JupyterLab, ``{{JupyterLab}} 3.0.15 documentation.'' Online:
  https://jupyterlab.readthedocs.io/en/stable/ (accessed 05.11.2021), 2021.

\bibitem{MPI4Py2005}
L.~Dalc{\'i}n, R.~Paz and M.~Storti, \emph{{{MPI}} for {{Python}}},
  \href{https://doi.org/10.1016/j.jpdc.2005.03.010}{\emph{Journal of Parallel
  and Distributed Computing} {\bfseries 65} (2005) 1108}.

\bibitem{MPI4Py2008}
L.~Dalc{\'i}n, R.~Paz, M.~Storti and J.~D'El{\'i}a, \emph{{{MPI}} for
  {{Python}}: {{Performance}} improvements and {{MPI}}-2 extensions},
  \href{https://doi.org/10.1016/j.jpdc.2007.09.005}{\emph{Journal of Parallel
  and Distributed Computing} {\bfseries 68} (2008) 655}.

\bibitem{Farzan:2017xzy}
Y.~Farzan and M.~Tortola, \emph{Neutrino oscillations and {{Non}}-{{Standard
  Interactions}}},
  \href{https://doi.org/10.3389/fphy.2018.00010}{\emph{Front.in Phys.}
  {\bfseries 6} (2018) 10} [\href{https://arxiv.org/abs/1710.09360}{{\ttfamily
  1710.09360}}].

\bibitem{Du:2021idh}
Y.~Du and J.-H. Yu, \emph{{Neutrino non-standard interactions meet precision
  measurements of N$_{eff}$}},
  \href{https://doi.org/10.1007/JHEP05(2021)058}{\emph{JHEP} {\bfseries 05}
  (2021) 058} [\href{https://arxiv.org/abs/2101.10475}{{\ttfamily
  2101.10475}}].

\bibitem{Stapleford:2016jgz}
C.~J. Stapleford, D.~J. V{\"a}{\"a}n{\"a}nen, J.~P. Kneller, G.~C. McLaughlin
  and B.~T. Shapiro, \emph{Nonstandard {{Neutrino Interactions}} in
  {{Supernovae}}},
  \href{https://doi.org/10.1103/PhysRevD.94.093007}{\emph{Phys.Rev.D}
  {\bfseries 94} (2016) } [\href{https://arxiv.org/abs/1605.04903}{{\ttfamily
  1605.04903}}].

\bibitem{AristizabalSierra:2018eqm}
D.~Aristizabal~Sierra, V.~De~Romeri and N.~Rojas, \emph{{{COHERENT}} analysis
  of neutrino generalized interactions},
  \href{https://doi.org/10.1103/PhysRevD.98.075018}{\emph{Phys.Rev.D}
  {\bfseries 98} (2018) } [\href{https://arxiv.org/abs/1806.07424}{{\ttfamily
  1806.07424}}].

\bibitem{Bischer:2019ttk}
I.~Bischer and W.~Rodejohann, \emph{General neutrino interactions from an
  effective field theory perspective},
  \href{https://doi.org/10.1016/j.nuclphysb.2019.114746}{\emph{Nucl.Phys.B}
  {\bfseries 947} (2019) } [\href{https://arxiv.org/abs/1905.08699}{{\ttfamily
  1905.08699}}].

\bibitem{Hoferichter:2020osn}
M.~Hoferichter, J.~Men{\'e}ndez and A.~Schwenk, \emph{Coherent elastic
  neutrino-nucleus scattering: {{EFT}} analysis and nuclear responses},
  \href{https://doi.org/10.1103/PhysRevD.102.074018}{\emph{Phys.Rev.D}
  {\bfseries 102} (2020) } [\href{https://arxiv.org/abs/2007.08529}{{\ttfamily
  2007.08529}}].

\bibitem{Healey:2013vka}
K.~J. Healey, A.~A. Petrov and D.~Zhuridov, \emph{{Nonstandard neutrino
  interactions and transition magnetic moments}},
  \href{https://doi.org/10.1103/PhysRevD.87.117301}{\emph{Phys. Rev. D}
  {\bfseries 87} (2013) 117301}
  [\href{https://arxiv.org/abs/1305.0584}{{\ttfamily 1305.0584}}].

\bibitem{Papoulias:2015iga}
D.~K. Papoulias and T.~S. Kosmas, \emph{Neutrino transition magnetic moments
  within the non-standard neutrino\textendash nucleus interactions},
  \href{https://doi.org/10.1016/j.physletb.2015.06.039}{\emph{Phys.Lett.B}
  {\bfseries 747} (2015) 454}
  [\href{https://arxiv.org/abs/1506.05406}{{\ttfamily 1506.05406}}].

\bibitem{Rohatgi2020}
A.~Rohatgi, ``Webplotdigitizer: Version 4.5.'' Online:
  https://automeris.io/WebPlotDigitizer (accessed 26.11.2021), 2021.

\bibitem{XENON:2020gfr}
{\scshape XENON} collaboration, E.~Aprile et~al., \emph{{Search for Coherent
  Elastic Scattering of Solar $^8$B Neutrinos in the XENON1T Dark Matter
  Experiment}},
  \href{https://doi.org/10.1103/PhysRevLett.126.091301}{\emph{Phys. Rev. Lett.}
  {\bfseries 126} (2021) 091301}
  [\href{https://arxiv.org/abs/2012.02846}{{\ttfamily 2012.02846}}].

\bibitem{Dorenbosch:1986tb}
{\scshape CHARM} collaboration, J.~Dorenbosch et~al., \emph{{Experimental
  Verification of the Universality of $\nu_e$ and $\nu_\mu$ Coupling to the
  Neutral Weak Current}},
  \href{https://doi.org/10.1016/0370-2693(86)90315-1}{\emph{Phys. Lett. B}
  {\bfseries 180} (1986) 303}.

\bibitem{Friedland:2011za}
A.~Friedland, M.~L. Graesser, I.~M. Shoemaker and L.~Vecchi, \emph{{Probing
  Nonstandard Standard Model Backgrounds with LHC Monojets}},
  \href{https://doi.org/10.1016/j.physletb.2012.06.078}{\emph{Phys. Lett. B}
  {\bfseries 714} (2012) 267}
  [\href{https://arxiv.org/abs/1111.5331}{{\ttfamily 1111.5331}}].

\bibitem{Coloma:2019mbs}
P.~Coloma, I.~Esteban, M.~C. Gonzalez-Garcia and M.~Maltoni, \emph{{Improved
  global fit to Non-Standard neutrino Interactions using COHERENT energy and
  timing data}}, \href{https://doi.org/10.1007/JHEP02(2020)023}{\emph{JHEP}
  {\bfseries 02} (2020) 023}
  [\href{https://arxiv.org/abs/1911.09109}{{\ttfamily 1911.09109}}].

\bibitem{Abdallah:2015ter}
J.~Abdallah, H.~Araujo, A.~Arbey, A.~Ashkenazi, A.~Belyaev, J.~Berger et~al.,
  \emph{Simplified {{Models}} for {{Dark Matter Searches}} at the {{LHC}}},
  \href{https://doi.org/10.1016/j.dark.2015.08.001}{\emph{Phys.Dark Univ.}
  {\bfseries 9-10} (2015) 8}
  [\href{https://arxiv.org/abs/1506.03116}{{\ttfamily 1506.03116}}].

\bibitem{Abercrombie:2015wmb}
D.~Abercrombie, N.~Akchurin, E.~Akilli, J.~Alcaraz~Maestre, B.~Allen,
  B.~Alvarez~Gonzalez et~al., \emph{Dark {{Matter Benchmark Models}} for
  {{Early LHC Run}}-2 {{Searches}}: {{Report}} of the {{ATLAS}}/{{CMS Dark
  Matter Forum}}},
  \href{https://doi.org/10.1016/j.dark.2019.100371}{\emph{Phys.Dark Univ.}
  {\bfseries 27} (2020) } [\href{https://arxiv.org/abs/1507.00966}{{\ttfamily
  1507.00966}}].

\bibitem{Boveia:2016mrp}
A.~Boveia et~al., \emph{{Recommendations on presenting LHC searches for missing
  transverse energy signals using simplified $s$-channel models of dark
  matter}}, \href{https://doi.org/10.1016/j.dark.2019.100365}{\emph{Phys. Dark
  Univ.} {\bfseries 27} (2020) 100365}
  [\href{https://arxiv.org/abs/1603.04156}{{\ttfamily 1603.04156}}].

\bibitem{Kahlhoefer:2015bea}
F.~Kahlhoefer, K.~Schmidt-Hoberg, T.~Schwetz and S.~Vogl, \emph{{Implications
  of unitarity and gauge invariance for simplified dark matter models}},
  \href{https://doi.org/10.1007/JHEP02(2016)016}{\emph{JHEP} {\bfseries 02}
  (2016) 016} [\href{https://arxiv.org/abs/1510.02110}{{\ttfamily
  1510.02110}}].

\bibitem{Ellis:2017tkh}
J.~Ellis, M.~Fairbairn and P.~Tunney, \emph{Anomaly-{{Free Dark Matter Models}}
  are not so {{Simple}}},
  \href{https://doi.org/10.1007/JHEP08(2017)053}{\emph{JHEP} {\bfseries 08}
  (2017) 053} [\href{https://arxiv.org/abs/1704.03850}{{\ttfamily
  1704.03850}}].

\bibitem{Morgante:2018tiq}
E.~Morgante, \emph{Simplified {{Dark Matter Models}}},
  \href{https://doi.org/10.1155/2018/5012043}{\emph{Adv.High Energy Phys.}
  {\bfseries 2018} (2018) } [\href{https://arxiv.org/abs/1804.01245}{{\ttfamily
  1804.01245}}].

\bibitem{Arcadi:2020gge}
G.~Arcadi, G.~Busoni, T.~Hugle and V.~T. Tenorth, \emph{{Comparing 2HDM $+$
  Scalar and Pseudoscalar Simplified Models at LHC}},
  \href{https://doi.org/10.1007/JHEP06(2020)098}{\emph{JHEP} {\bfseries 06}
  (2020) 098} [\href{https://arxiv.org/abs/2001.10540}{{\ttfamily
  2001.10540}}].

\bibitem{Vignaroli:2019lkg}
N.~Vignaroli, \emph{{Leptoquarks in $B$-meson anomalies: simplified models and
  HL-LHC discovery prospects}},
  \href{https://doi.org/10.1393/ncc/i2020-20053-0}{\emph{Nuovo Cim. C}
  {\bfseries 43} (2020) 53} [\href{https://arxiv.org/abs/1912.00899}{{\ttfamily
  1912.00899}}].

\bibitem{Bauer:2018onh}
M.~Bauer, P.~Foldenauer and J.~Jaeckel, \emph{Hunting {{All}} the {{Hidden
  Photons}}}, \href{https://doi.org/10.1007/JHEP07(2018)094}{\emph{JHEP}
  {\bfseries 07} (2018) 094}
  [\href{https://arxiv.org/abs/1803.05466}{{\ttfamily 1803.05466}}].

\bibitem{Arcadi:2017hfi}
G.~Arcadi, M.~D. Campos, M.~Lindner, A.~Masiero and F.~S. Queiroz, \emph{{Dark
  sequential Z' portal: Collider and direct detection experiments}},
  \href{https://doi.org/10.1103/PhysRevD.97.043009}{\emph{Phys. Rev. D}
  {\bfseries 97} (2018) 043009}
  [\href{https://arxiv.org/abs/1708.00890}{{\ttfamily 1708.00890}}].

\bibitem{Denton:2018xmq}
P.~B. Denton, Y.~Farzan and I.~M. Shoemaker, \emph{Testing large non-standard
  neutrino interactions with arbitrary mediator mass after {{COHERENT}} data},
  \href{https://doi.org/10.1007/JHEP07(2018)037}{\emph{JHEP} {\bfseries 07}
  (2018) 037} [\href{https://arxiv.org/abs/1804.03660}{{\ttfamily
  1804.03660}}].

\bibitem{Aaboud:2016cth}
{\scshape ATLAS} collaboration, M.~Aaboud et~al., \emph{{Search for high-mass
  new phenomena in the dilepton final state using proton-proton collisions at
  $\sqrt{s}=13$ TeV with the ATLAS detector}},
  \href{https://doi.org/10.1016/j.physletb.2016.08.055}{\emph{Phys. Lett. B}
  {\bfseries 761} (2016) 372}
  [\href{https://arxiv.org/abs/1607.03669}{{\ttfamily 1607.03669}}].

\bibitem{Batell:2009di}
B.~Batell, M.~Pospelov and A.~Ritz, \emph{Exploring {{Portals}} to a {{Hidden
  Sector Through Fixed Targets}}},
  \href{https://doi.org/10.1103/PhysRevD.80.095024}{\emph{Phys.Rev.D}
  {\bfseries 80} (2009) } [\href{https://arxiv.org/abs/0906.5614}{{\ttfamily
  0906.5614}}].

\bibitem{Bjorken:2009mm}
J.~D. Bjorken, R.~Essig, P.~Schuster and N.~Toro, \emph{New {{Fixed}}-{{Target
  Experiments}} to {{Search}} for {{Dark Gauge Forces}}},
  \href{https://doi.org/10.1103/PhysRevD.80.075018}{\emph{Phys.Rev.D}
  {\bfseries 80} (2009) } [\href{https://arxiv.org/abs/0906.0580}{{\ttfamily
  0906.0580}}].

\bibitem{Bilmis:2015lja}
S.~Bilmis, I.~Turan, T.~M. Aliev, M.~Deniz, L.~Singh and H.~T. Wong,
  \emph{Constraints on {{Dark Photon}} from {{Neutrino}}-{{Electron Scattering
  Experiments}}},
  \href{https://doi.org/10.1103/PhysRevD.92.033009}{\emph{Phys.Rev.D}
  {\bfseries 92} (2015) } [\href{https://arxiv.org/abs/1502.07763}{{\ttfamily
  1502.07763}}].

\bibitem{Lindner:2018kjo}
M.~Lindner, F.~S. Queiroz, W.~Rodejohann and X.-J. Xu, \emph{{Neutrino-electron
  scattering: general constraints on Z' and dark photon models}},
  \href{https://doi.org/10.1007/JHEP05(2018)098}{\emph{JHEP} {\bfseries 05}
  (2018) 098} [\href{https://arxiv.org/abs/1803.00060}{{\ttfamily
  1803.00060}}].

\bibitem{Lees:2014xha}
{\scshape BaBar} collaboration, J.~P. Lees et~al., \emph{{Search for a Dark
  Photon in $e^+e^-$ Collisions at BaBar}},
  \href{https://doi.org/10.1103/PhysRevLett.113.201801}{\emph{Phys. Rev. Lett.}
  {\bfseries 113} (2014) 201801}
  [\href{https://arxiv.org/abs/1406.2980}{{\ttfamily 1406.2980}}].

\bibitem{Lees:2017lec}
{\scshape BaBar} collaboration, J.~P. Lees et~al., \emph{{Search for Invisible
  Decays of a Dark Photon Produced in ${e}^{+}{e}^{-}$ Collisions at BaBar}},
  \href{https://doi.org/10.1103/PhysRevLett.119.131804}{\emph{Phys. Rev. Lett.}
  {\bfseries 119} (2017) 131804}
  [\href{https://arxiv.org/abs/1702.03327}{{\ttfamily 1702.03327}}].

\bibitem{Aaij:2017rft}
{\scshape LHCb} collaboration, R.~Aaij et~al., \emph{{Search for Dark Photons
  Produced in 13 TeV $pp$ Collisions}},
  \href{https://doi.org/10.1103/PhysRevLett.120.061801}{\emph{Phys. Rev. Lett.}
  {\bfseries 120} (2018) 061801}
  [\href{https://arxiv.org/abs/1710.02867}{{\ttfamily 1710.02867}}].

\bibitem{Ilten:2018crw}
P.~Ilten, Y.~Soreq, M.~Williams and W.~Xue, \emph{Serendipity in dark photon
  searches}, \href{https://doi.org/10.1007/JHEP06(2018)004}{\emph{JHEP}
  {\bfseries 06} (2018) 004}
  [\href{https://arxiv.org/abs/1801.04847}{{\ttfamily 1801.04847}}].

\bibitem{Harnik:2012ni}
R.~Harnik, J.~Kopp and P.~A.~N. Machado, \emph{{Exploring nu Signals in Dark
  Matter Detectors}},
  \href{https://doi.org/10.1088/1475-7516/2012/07/026}{\emph{JCAP} {\bfseries
  07} (2012) 026} [\href{https://arxiv.org/abs/1202.6073}{{\ttfamily
  1202.6073}}].

\bibitem{Miranda:2020tif}
O.~G. Miranda, D.~K. Papoulias, G.~Sanchez~Garcia, O.~Sanders, M.~T\'ortola and
  J.~W.~F. Valle, \emph{{Implications of the first detection of coherent
  elastic neutrino-nucleus scattering (CEvNS) with Liquid Argon}},
  \href{https://doi.org/10.1007/JHEP05(2020)130}{\emph{JHEP} {\bfseries 05}
  (2020) 130} [\href{https://arxiv.org/abs/2003.12050}{{\ttfamily
  2003.12050}}].

\bibitem{Bell:2016ekl}
N.~F. Bell, G.~Busoni and I.~W. Sanderson, \emph{{Self-consistent Dark Matter
  Simplified Models with an s-channel scalar mediator}},
  \href{https://doi.org/10.1088/1475-7516/2017/03/015}{\emph{JCAP} {\bfseries
  03} (2017) 015} [\href{https://arxiv.org/abs/1612.03475}{{\ttfamily
  1612.03475}}].

\bibitem{DelNobile:2013sia}
M.~Cirelli, E.~Del~Nobile and P.~Panci, \emph{Tools for model-independent
  bounds in direct dark matter searches},
  \href{https://doi.org/10.1088/1475-7516/2013/10/019}{\emph{JCAP} {\bfseries
  10} (2013) 019} [\href{https://arxiv.org/abs/1307.5955}{{\ttfamily
  1307.5955}}].

\bibitem{CONUSquenching}
A.~Bonhomme et~al., \emph{{Direct measurement of the ionization quenching
  factor of nuclear recoils in germanium in the keV energy range}},
  \href{https://arxiv.org/abs/2202.03754}{{\ttfamily 2202.03754}}.

\bibitem{Blennow:2019bvl}
M.~Blennow, E.~Fernandez-Martinez, T.~Ota and S.~Rosauro-Alcaraz,
  \emph{{Physics potential of the ESS$\nu$SB}},
  \href{https://doi.org/10.1140/epjc/s10052-020-7761-9}{\emph{Eur. Phys. J. C}
  {\bfseries 80} (2020) 190}
  [\href{https://arxiv.org/abs/1912.04309}{{\ttfamily 1912.04309}}].

\bibitem{Baxter:2019mcx}
D.~Baxter, J.~I. Collar, P.~Coloma, C.~E. Dahl, I.~Esteban, P.~Ferrario et~al.,
  \emph{Coherent {{Elastic Neutrino}}-{{Nucleus Scattering}} at the {{European
  Spallation Source}}},
  \href{https://doi.org/10.1007/JHEP02(2020)123}{\emph{JHEP} {\bfseries 02}
  (2020) 123} [\href{https://arxiv.org/abs/1911.00762}{{\ttfamily
  1911.00762}}].

\bibitem{Dent:2017mpr}
J.~B. Dent, B.~Dutta, S.~Liao, J.~L. Newstead, L.~E. Strigari and J.~W. Walker,
  \emph{Accelerator and reactor complementarity in coherent neutrino-nucleus
  scattering},
  \href{https://doi.org/10.1103/PhysRevD.97.035009}{\emph{Phys.Rev.D}
  {\bfseries 97} (2018) } [\href{https://arxiv.org/abs/1711.03521}{{\ttfamily
  1711.03521}}].

\bibitem{Lindner:2017nmb}
M.~Lindner, W.~Rodejohann and X.-J. Xu, \emph{Neutrino {{Parameters}} from
  {{Reactor}} and {{Accelerator Neutrino Experiments}}},
  \href{https://doi.org/10.1103/PhysRevD.97.075024}{\emph{Phys.Rev.D}
  {\bfseries 97} (2018) } [\href{https://arxiv.org/abs/1709.10252}{{\ttfamily
  1709.10252}}].

\bibitem{Bellenghi:2019vtc}
C.~Bellenghi, D.~Chiesa, L.~Di~Noto, M.~Pallavicini, E.~Previtali and
  M.~Vignati, \emph{{Coherent elastic nuclear scattering of $^{51}$Cr
  neutrinos}}, \href{https://doi.org/10.1140/epjc/s10052-019-7240-3}{\emph{Eur.
  Phys. J. C} {\bfseries 79} (2019) 727}
  [\href{https://arxiv.org/abs/1905.10611}{{\ttfamily 1905.10611}}].

\bibitem{Cadeddu:2019qmv}
M.~Cadeddu, F.~Dordei, C.~Giunti, K.~A. Kouzakov, E.~Picciau and A.~I.
  Studenikin, \emph{{Potentialities of a low-energy detector based on $^4$He
  evaporation to observe atomic effects in coherent neutrino scattering and
  physics perspectives}},
  \href{https://doi.org/10.1103/PhysRevD.100.073014}{\emph{Phys. Rev. D}
  {\bfseries 100} (2019) 073014}
  [\href{https://arxiv.org/abs/1907.03302}{{\ttfamily 1907.03302}}].

\bibitem{Abdullah:2020iiv}
M.~Abdullah, D.~Aristizabal~Sierra, B.~Dutta and L.~E. Strigari, \emph{Coherent
  {{Elastic Neutrino}}-{{Nucleus Scattering}} with directional detectors},
  \href{https://doi.org/10.1103/PhysRevD.102.015009}{\emph{Phys.Rev.D}
  {\bfseries 102} (2020) } [\href{https://arxiv.org/abs/2003.11510}{{\ttfamily
  2003.11510}}].

\bibitem{AristizabalSierra:2021uob}
D.~Aristizabal~Sierra, B.~Dutta, D.~Kim, D.~Snowden-Ifft and L.~E. Strigari,
  \emph{{Coherent elastic neutrino-nucleus scattering with the
  \ensuremath{\nu}BDX-DRIFT directional detector at next generation neutrino
  facilities}}, \href{https://doi.org/10.1103/PhysRevD.104.033004}{\emph{Phys.
  Rev. D} {\bfseries 104} (2021) 033004}
  [\href{https://arxiv.org/abs/2103.10857}{{\ttfamily 2103.10857}}].

\bibitem{Tomalak:2020zfh}
O.~Tomalak, P.~Machado, V.~Pandey and R.~Plestid, \emph{Flavor-dependent
  radiative corrections in coherent elastic neutrino-nucleus scattering},
  \href{https://doi.org/10.1007/JHEP02(2021)097}{\emph{JHEP} {\bfseries 02}
  (2021) 097} [\href{https://arxiv.org/abs/2011.05960}{{\ttfamily
  2011.05960}}].

\end{thebibliography}\endgroup
\bibliographystyle{JHEP}

\end{document}